\documentclass[aps,prd,preprint,groupedaddress]{revtex4}
\usepackage{hyperref}
\usepackage{amssymb}
\usepackage{amsmath}
\usepackage{slashed}
\usepackage{graphicx}
\usepackage{color}

\newcommand{\ohf}{\ensuremath{\textstyle \frac{1}{2}}}
\newcommand{\thf}{\ensuremath{\textstyle \frac{3}{2}}}
\newcommand{\fhf}{\ensuremath{\textstyle \frac{5}{2}}}
\newcommand{\shf}{\ensuremath{\textstyle \frac{7}{2}}}
\newcommand{\otd}{\ensuremath{\textstyle \frac{1}{3}}}
\newcommand{\osx}{\ensuremath{\textstyle \frac{1}{6}}}

\newcommand{\rthreeotwo}{\ensuremath{\textstyle \frac{\sqrt 3}{2}}}

\newcommand{\gts}{\ensuremath{\mr{GTS}}}
\newcommand{\sugts}{\ensuremath{SU(3)_F\times\mr{GTS}}}
\newcommand{\isosgts}{\ensuremath{SU(2)_I\times U(1)_z\times\mr{GTS}}}
\newcommand{\spinks}{\ensuremath{SU(2)_S\times SU(12)_f}}
\newcommand{\su}[2]{\ensuremath{SU(#1)_{#2}}}

\newcommand{\mc}[1]{\ensuremath{\mathcal{#1}}}
\newcommand{\mr}[1]{\ensuremath{\mathrm{#1}}}
\newcommand{\mb}[1]{\ensuremath{\mathbf{#1}}}

\newcommand{\nsig}{\ensuremath{\Sigma^0}}
\newcommand{\lam}{\ensuremath{\Lambda}}

\begin{document}

\title{Staggered Baryon Operators with Flavor $SU(3)$ Quantum Numbers}

\author{Jon A. Bailey}
\email[]{jabailey@wustl.edu}
\affiliation{Washington University in St. Louis}

\date{\today}

\begin{abstract}
The construction of the first baryon operators for staggered lattice QCD exploited the taste symmetry to emulate physical quark flavor; contemporary 2+1 flavor simulations explicitly include three physical quark flavors and necessitate interpreting a valence sector with twelve quarks.  After discussing expected features of the resulting baryon spectrum, I consider the spectra of operators transforming irreducibly under $SU(3)_F\times\mathrm{GTS}$, the direct product of flavor $SU(3)_F$ and the geometrical time-slice group of the 1-flavor staggered theory.  I then describe the construction of a set of maximally local baryon operators transforming irreducibly under $SU(3)_F\times\mathrm{GTS}$ and enumerate this set.  In principle, the operators listed here could be used to extract the masses of all the lightest spin-$\frac{1}{2}$ and spin-$\frac{3}{2}$ baryon resonances of staggered QCD.  Using appropriate operators from this set in partially quenched simulations should allow for particularly clean 2+1 flavor calculations of the masses of the nucleon, $\Delta$, $\Sigma^*$, $\Xi^*$, and $\Omega^-$.
\end{abstract}


\maketitle

\section{\label{intro}Introduction}
The advantages \cite{Toussaint:2001zc} of the staggered formulation \cite{Kogut:1974ag} of lattice QCD have allowed successful calculations of many hadron masses, decay constants, and other quantities of phenomenological importance~\cite{Aubin:2004ck,Aubin:2004ej,Aubin:2004fs,Aubin:2004wf,Aubin:2005ar,Davies:2003ik}.  However, to date this success has been most pronounced in the meson sector.  Progress calculating baryonic quantities has been impeded by comparatively large statistical errors, questions of interpretation affecting systematics, and a lack of staggered chiral forms, which would improve control over the chiral and continuum limits.  

Moreover, the staggered formulation incorporates the conjecture~\cite{Marinari:1981qf} that taking the fourth root of the fermion determinant eliminates, in the continuum limit, the effects of remnant quark doublers in the sea.  If we are to know that the rooted staggered formulation is an authentic regulator of QCD, this conjecture must be demonstrated.  The last few months have seen good progress in our understanding of the fourth root trick~\cite{Bernard:2006ee,Shamir:2006nj} and the replica trick \cite{Aubin:2003rg,Aubin:2003mg,Aubin:2003uc} used to incorporate the fourth root in staggered chiral perturbation theory~\cite{Bernard:2006zw}.  However, the results of \cite{Bernard:2006zw} have yet to be extended to the baryon sector, and many of the assumptions of \cite{Bernard:2006zw,Shamir:2006nj} have not yet been tested.  Lacking proof of these assumptions, calculations of experimentally well-known baryonic quantities provide valuable additional tests of the validity of the fourth root and replica tricks.  With an eye toward such calculations, this paper addresses the questions of interpretation encountered in 2+1 flavor staggered simulations of the lightest octet and decuplet baryons.  

These questions were first considered in~\cite{Golterman:1984dn}.  In that work baryon operators transforming within irreducible representations (irreps) of the lattice symmetry group of a time slice (the ``geometrical time-slice group,'' \mr{GTS}) were constructed from elementary staggered quark fields.  The four degenerate pseudoflavors of the staggered formulation were interpreted as physical quark flavors with the idea that one could introduce additional terms in the action to break the degeneracy~\cite{Golterman:1984cy,Golterman:1984dn,Golterman:1984ds}.  This approach proved to involve fine-tuning of the symmetry-breaking terms and other difficulties~\cite{Golterman:1984ds,Sharpe:2006re} and has since fallen into disfavor.

In contemporary studies employing staggered fermions, three physical quark flavors are explicitly included in the action~\cite{Aubin:2004wf}; the four pseudoflavors per physical quark flavor are called ``tastes.''  From this perspective, the interpolating fields of~\cite{Golterman:1984dn} correspond to staggered QCD with only one physical quark flavor.  However, one can still use the operators of~\cite{Golterman:1984dn} to perform simulations of baryons with nondegenerate valence quarks:  Varying the quark masses in the propagators amounts to specifying interpolating fields that have definite upness, downness, and strangeness.  Unfortunately, operators lacking definite isospin do not distinguish the \nsig\ from the \lam\ and cannot be used to study isospin breaking.  Operators transforming within irreps of \su{3}{F} accommodate isospin automatically, so one might hope to use such operators to address these issues directly.

When working with staggered fermions, operators possessing \su{3}{F} quantum numbers confer advantages that operators possessing only isospin cannot confer.  The presence of taste quantum numbers implies that the valence sector of staggered QCD contains more baryons than nature does.  The restoration of taste symmetry in the continuum limit plays a key role in~\cite{Bernard:2006ee,Shamir:2006nj,Bernard:2006zw}, and I assume taste restoration throughout.  Then the staggered baryons are degenerate, in the continuum limit, within irreps of the taste symmetry group, \su{4}{T}, and baryons composed of valence quarks of the same taste are easily seen to be degenerate with physical states.  At nonzero lattice spacing, discretization effects lift the degeneracies and induce mixing among baryons with different \su{4}{T} quantum numbers.  The resulting spectrum consists of many nearly degenerate states separated by energies crudely estimated to be 10-40 MeV~\cite{Bailey:2006ua}.  Although presenting no problem of principle, this situation makes spectrum calculations difficult.  However, for calculating the isospin limits of the masses of the nucleon and the lightest decuplet baryons, I show in Sec.'s~\ref{spec} and~\ref{operspec} that a judicious choice of quark masses and operators transforming within irreps of \sugts\ allows one to interpolate to states that are not part of nearly degenerate multiplets, thus avoiding the practical difficulties presented by the splittings and mixings in the staggered baryon spectrum.

In Sec.~\ref{spec}, I note two basic assumptions about the baryon mass spectrum of staggered QCD, present the continuum and lattice symmetries of the valence sector of staggered QCD, and discuss the implied features of the mass spectrum of the lightest spin-\ohf\ and spin-\thf\ staggered baryons.  Given these features one can deduce semi-quantitative information about the spectra of operators transforming irreducibly under \sugts.  This information in turn allows one to make specific statements about the utility and limitations of such operators; Sec.~\ref{operspec} contains a discussion of the spectra of operators transforming irreducibly under \sugts\ and the implications for practical spectrum calculations.  In Sec.~\ref{cons}, I extend the approach of~\cite{Golterman:1984dn} to construct operators transforming within irreps of \sugts.  Section~\ref{conclu} contains a summary and some comments about related work.  Couplings of the operators to excited states are discussed in an appendix.

\section{\label{spec}Symmetry and the staggered baryon spectrum}
In the $SU(6)$ quark model, the lightest octet and decuplet baryons transform within spin-flavor $SU(2)_S\times SU(3)_F$ irreps embedded in the symmetric irrep of $SU(6)$,
\begin{eqnarray}
SU(6)&\supset& SU(2)_S\times SU(3)_F\nonumber\\
\mb{56_S}&\rightarrow& \mb{(\ohf,\ 8_M)}\oplus\mb{(\thf,\ 10_S)},\label{su6}
\end{eqnarray}
where the subscripts indicate the symmetry of each irrep, either symmetric or mixed.  Independently of whether one takes the fourth root in the sea, the valence sector of staggered QCD with three light quarks possesses an $SU(12)_f$ flavor-taste symmetry; assuming that the staggered baryons fall into the symmetric and mixed three-quark irreps of $SU(2)_S\times SU(12)_f$ embedded in the symmetric irrep of $SU(24)$ means considering 
\begin{eqnarray}
SU(24)&\supset&SU(2)_S\times SU(12)_f \nonumber\\
\mb{2600_S}&\rightarrow& \mb{(\ohf,\ 572_M)}\oplus\mb{(\thf,\ 364_S)}.\label{symmetric2600}
\end{eqnarray}
To deduce information about the masses of these baryons, I further assume that baryons constructed out of valence quarks of the same taste (``single-taste baryons'') become degenerate, in the continuum limit, with physical baryons.  This assumption must be true if staggered QCD describes the real world:  In the continuum limit, taste restoration implies that each quark flavor comes in four equivalent tastes.  If we choose the same taste for each flavor, then the taste degree of freedom is irrelevant, and baryons constructed in accord with~(\ref{symmetric2600}) necessarily possess the same spin and flavor structure as the physical baryons of~(\ref{su6}).  

The first step in the analysis is to disentangle the flavor and taste quantum numbers.  Decomposing the $SU(12)_f$ irreps into irreps of the direct product of flavor $SU(3)_F$ and taste $SU(4)_T$ gives
\begin{subequations}
\label{ks2ft}
\begin{eqnarray}
SU(12)_f&\supset& SU(3)_F\times SU(4)_T\nonumber \\
\mb{572_M}&\rightarrow&\mb{(10_S,\ 20_M)\oplus(8_M,\ 20_S)\oplus(8_M,\ 20_M)\oplus(8_M,\ \bar 4_A)\oplus(1_A,\ 20_M)}\label{572ft}\\
\mb{364_S}&\rightarrow&\mb{(10_S,\ 20_S)\oplus(8_M,\ 20_M)\oplus(1_A,\ \bar 4_A)},\label{364ft}
\end{eqnarray}
\end{subequations}
where the symmetries of the irreps are symmetric, mixed, or antisymmetric.  Single-taste baryons must transform in the symmetric three-quark irrep of \su{4}{T}, the \mb{20_S}; this irrep contains precisely four single-taste members, one for each quark taste.  Examining the decompositions~(\ref{ks2ft}), we see that only the \mb{(8_M,\ 20_S)} and the \mb{(10_S,\ 20_S)} contain single-taste baryons.  As expected if these baryons are to be degenerate with the physical baryons, there exist a spin-\ohf\ octet and a spin-\thf\ decuplet for each of the four valence quark tastes.  In the continuum limit, members within the same \su{4}{T} irrep are degenerate.  Therefore we expect all 20 octets of the \mb{(8_M,\ 20_S)} to become degenerate with the physical octet, and all 20 decuplets in the \mb{(10_S,\ 20_S)}, with the lightest decuplet.

To make further progress, we consider the continuum flavor symmetries as well as the taste symmetry.  In what follows, $m_x$, $m_y$, and $m_z$ respectively denote the masses of the up, down, and strange valence quarks, $m_u$, $m_d$, and $m_s$ denote the physical quark masses, and $\hat m=\ohf(m_u+m_d)$.  There are four cases of interest, which are summarized in Table~\ref{cases}.  The second line of Table~\ref{cases} corresponds to fully dynamical, 2+1 flavor QCD, while the other cases turn out to be useful, when used with certain operators identified in Sec.~\ref{operspec}, for cleanly extracting the masses of physical states.  For clarity, the valence masses and physical masses are taken to be related as shown in Table~\ref{cases} and the masses of the up, down, and strange sea quarks are respectively set equal to $\hat m$, $\hat m$, and $m_s$ throughout.  However, the valence symmetries depend only on the lattice spacing and the relative values of the valence quark masses; accordingly, conclusions about the existence of degeneracies and mixings, which depend upon the valence symmetries alone, hold for all simulations in which the valence masses are related as shown in Table~\ref{cases}.  Since we are concerned with the consequences of the valence sector symmetries, we use ``isospin'' and ``strangeness'' to refer to quantum numbers of valence quarks.  Finally, until further notice, we work in the continuum limit.  
\begin{table}
\begin{ruledtabular}
\begin{tabular}{lll}
Case & Symmetry for $a=0$ & Symmetry for $a\neq 0$ \\ \hline
$m_x=m_y=m_z=\hat m$ & $SU(12)_f$ & \sugts \\
$m_x=m_y=\hat m,\;m_z=m_s$ & $SU(8)_{x,y}\times SU(4)_z$ & \isosgts \\
$m_x=m_y=m_z=m_s$ & $SU(12)_f$ & \sugts \\
$m_x=m_y=m_s,\;m_z=\hat m$ & $SU(8)_{x,y}\times SU(4)_z$ & \isosgts 
\end{tabular}
\end{ruledtabular}
\caption{\label{cases}The valence quark masses and the valence sector symmetries for the four cases of interest; $\hat m=\ohf(m_u+m_d)$, and $m_u$, $m_d$, and $m_s$ denote the physical quark masses.}
\end{table}

Setting $m_x=m_y=m_z=\hat m$, the valence sector possesses the full $SU(12)_f$ symmetry, and the baryons are degenerate within irreps of $SU(2)_S\times SU(12)_f$.  All members of the \mb{572_M} are degenerate with the single-taste nucleon (of the \mb{(8_M,\ 20_S)}), and all members of the \mb{364_S}, with the single-taste $\Delta$ (of the \mb{(10_S,\ 20_S)}).  If we now break flavor $SU(3)_F$ by taking $m_z=m_s$, the spin-\ohf\ and spin-\thf\ baryons with zero strangeness remain degenerate, respectively, with the nucleon or the $\Delta$, while baryons with nonzero strangeness have masses that reflect the remaining valence sector symmetry.  

We can freely rotate the eight up and down valence quarks together or, independently, the four strange valence quarks.  Therefore the valence symmetry for the second case in Table~\ref{cases} is $SU(8)_{x,y}\times SU(4)_z$; the taste $SU(4)_T$ symmetry is the subgroup of this group consisting of identical $SU(4)$ transformations on all three quark flavors.  The members of the \mb{572_M} and \mb{364_S} are degenerate within irreps of $SU(8)_{x,y}\times SU(4)_z$.  To identify the baryons in the \mb{572_M} and the \mb{364_S} that are degenerate with the octet and decuplet baryons, we perform a series of decompositions:  \begin{eqnarray*}
\su{12}{f}&\supset& SU(8)_{x,y}\times SU(4)_z\\
SU(8)_{x,y}\times SU(4)_z&\supset& SU(2)_I\times\su{4}{T}\\
\su{3}{F}\times\su{4}{T}&\supset&\su{2}{I}\times\su{4}{T}.
\end{eqnarray*}
The first decomposition reveals the $SU(8)_{x,y}\times SU(4)_z$ irreps in which the baryons are degenerate.  The second decomposition allows us to identify the $\su{2}{I}\times U(1)_z\times\su{4}{T}$ irreps of the various members of the $SU(8)_{x,y}\times SU(4)_z$ irreps; as we will see below, the strangeness is revealed by the \su{4}{z} irrep.  The last decomposition connects the results of the first two with the more physically transparent decompositions~(\ref{ks2ft}); the presence of an \su{4}{T} \mb{20_S} (taste-symmetric) irrep alerts us to the presence of baryons that have physical masses, while the degeneracies are dictated by $SU(8)_{x,y}\times SU(4)_z$.  Explicitly,
\begin{subequations}
\label{ks2su84}
\begin{eqnarray}
SU(12)_f&\supset& SU(8)_{x,y}\times SU(4)_z\nonumber \\
\mb{572_M}&\rightarrow&\mb{(168_M,\ 1)\oplus(28_A,\ 4)\oplus(36_S,\ 4)}\nonumber \\
&&\oplus\;\mb{(8,\ 6_A)\oplus(8,\ 10_S)\oplus(1,\ 20_M)}\label{572su84}\\
\mb{364_S}&\rightarrow&\mb{(120_S,\ 1)\oplus(36_S,\ 4)\oplus(8,\ 10_S)\oplus(1,\ 20_S)},\label{364su84}
\end{eqnarray}
\end{subequations}
\begin{subequations}
\label{su842isot}
\begin{eqnarray}
SU(8)_{x,y}\times SU(4)_z&\supset& SU(2)_I\times SU(4)_T\nonumber \\
\mb{(168_M,\ 1)}_0&\rightarrow&\mb{(\thf,\ 20_M)}_0\oplus\mb{(\ohf,\ 20_S)}_0\oplus\mb{(\ohf,\ \bar 4_A)}_0\oplus\mb{(\ohf,\ 20_M)}_0\label{1682isot}\\
\mb{(28_A,\ 4)}_{-1}&\rightarrow &\mb{(1,\ 20_M)}_{-1}\oplus\mb{(1,\ \bar 4_A)}_{-1}\oplus\mb{(0,\ 20_S)}_{-1}\oplus\mb{(0,\ 20_M)}_{-1}\label{282isot}\\
\mb{(36_S,\ 4)}_{-1}&\rightarrow &\mb{(1,\ 20_S)}_{-1}\oplus\mb{(1,\ 20_M)}_{-1}\oplus\mb{(0,\ 20_M)}_{-1}\oplus\mb{(0,\ \bar 4_A)}_{-1}\label{362isot}\\
\mb{(8,\ 6_A)}_{-2}&\rightarrow &\mb{(\ohf,\ 20_M)}_{-2}\oplus\mb{(\ohf,\ \bar 4_A)}_{-2}\label{862isot}\\
\mb{(8,\ 10_S)}_{-2}&\rightarrow &\mb{(\ohf,\ 20_M)}_{-2}\oplus\mb{(\ohf,\ 20_S)}_{-2}\label{8102isot}\\
\mb{(1,\ 20_M)}_{-3}&\rightarrow &\mb{(0,\ 20_M)}_{-3}\label{202isot}\\
\mb{(120_S,\ 1)}_0 &\rightarrow &\mb{(\thf,\ 20_S)}_{0}\oplus\mb{(\ohf,\ 20_M)}_0 \label{1202isot}\\
\mb{(1,\ 20_S)}_{-3}&\rightarrow&\mb{(0,\ 20_S)}_{-3},\label{0202isot}
\end{eqnarray}
\end{subequations}
where all numbers in parentheses denote the dimensions of the corresponding irreps except those listed for \su{2}{I}, where the total isospin quantum number is given; subscripts outside the parentheses indicate the strangeness of the irrep.  The strangeness of the $SU(8)_{x,y}\times SU(4)_z$ irreps can be deduced by considering the transformations of the members of these irreps under $SU(4)_z$.  For example, members of the \mb{(168_M,\ 1)} of $SU(8)_{x,y}\times SU(4)_z$ are invariant under $SU(4)_z$; therefore, they have no valence strangeness:  $Z=0$.  Members of the \mb{(28_A,\ 4)} and \mb{(36_S,\ 4)} transform in the fundamental representation (rep) of $SU(4)_z$, and therefore have $Z=-1$.  The \mb{6_A} and \mb{10_S} are the antisymmetric and symmetric linear combinations of two fundamental reps of $SU(4)$, so these irreps carry $Z=-2$.  

To connect these results with the decompositions~(\ref{ks2ft}), consider the decomposition of the $\su{3}{F}\times\su{4}{T}$ irreps in (\ref{ks2ft}) into irreps of $\su{2}{I}\times\su{4}{T}$:
\begin{subequations}
\label{ft2isot}
\begin{eqnarray}
SU(3)_F\times SU(4)_T&\supset& SU(2)_I\times SU(4)_T\nonumber \\
\mb{(10_S,\ 20_M)}&\rightarrow&\mb{(\thf,\ 20_M)}_0\oplus\mb{(1,\ 20_M)}_{-1}\oplus\mb{(\ohf,\ 20_M)}_{-2}\oplus\mb{(0,\ 20_M)}_{-3}\label{ft1020m2isot}\\
\mb{(8_M,\ 20_S)}&\rightarrow&\mb{(\ohf,\ 20_S)}_{0}\oplus\mb{(1,\ 20_S)}_{-1}\oplus\mb{(0,\ 20_S)}_{-1}\oplus\mb{(\ohf,\ 20_S)}_{-2}\label{ft820s2isot}\\
\mb{(8_M,\ 20_M)}&\rightarrow&\mb{(\ohf,\ 20_M)}_{0}\oplus\mb{(1,\ 20_M)}_{-1}\oplus\mb{(0,\ 20_M)}_{-1}\oplus\mb{(\ohf,\ 20_M)}_{-2}\label{ft820m2isot}\\
\mb{(8_M,\ \bar 4_A)}&\rightarrow&\mb{(\ohf,\ \bar 4_A)}_{0}\oplus\mb{(1,\ \bar 4_A)}_{-1}\oplus\mb{(0,\ \bar 4_A)}_{-1}\oplus\mb{(\ohf,\ \bar 4_A)}_{-2}\label{ft84a2isot}\\
\mb{(1_A,\ 20_M)}&\rightarrow&\mb{(0,\ 20_M)}_{-1}\label{1a20m2isot}\\
\mb{(10_S,\ 20_S)}&\rightarrow&\mb{(\thf,\ 20_S)}_0\oplus\mb{(1,\ 20_S)}_{-1}\oplus\mb{(\ohf,\ 20_S)}_{-2}\oplus\mb{(0,\ 20_S)}_{-3}\label{ft1020s2isot}\\
\mb{(1_A,\ \bar 4_A)}&\rightarrow&\mb{(0,\ \bar 4_A)}_{-1}.\label{1a4a2isot}
\end{eqnarray}
\end{subequations}
Substituting these results into (\ref{ks2ft}) gives the decomposition of the $SU(12)_f$ irreps under $\su{2}{I}\times \su{4}{T}$.  The same decompositions must result if one substitutes the decompositions~(\ref{su842isot}) into the decompositions~(\ref{ks2su84}); this fact furnishes a nice consistency check of the decompositions.  

Consider the $SU(8)_{x,y}\times SU(4)_z$ irreps appearing in~(\ref{364su84}), the decomposition of the \mb{364_S}.  Comparing their decompositions~(\ref{1202isot}), (\ref{362isot}), (\ref{8102isot}), and (\ref{0202isot}) with the decomposition given in~(\ref{ft1020s2isot}), we note that the taste-symmetric irreps correspond one-to-one.  Now the staggered baryons are degenerate within irreps of $SU(8)_{x,y}\times SU(4)_z$, and we are assuming that the single-taste baryons have physical masses.  Therefore, all members of the \mb{(120_S,\ 1)} are degenerate with the $\Delta$, members of the \mb{(36_S,\ 4)} appearing in~(\ref{364su84}) are degenerate with the $\Sigma^*$, members of the \mb{(8,\ 10_S)} in~(\ref{364su84}) are degenerate with the $\Xi^*$, and members of the \mb{(1,\ 20_S)}, with the $\Omega^-$.  Noting the strangeness of these $SU(8)_{x,y}\times SU(4)_z$ irreps, we see that all members of the \mb{364_S} having the same strangeness are degenerate, and all baryons of the \mb{(\thf,\ 364_S)} have masses that are degenerate with the baryons of the lightest physical decuplet.  

Now consider the $SU(8)_{x,y}\times SU(4)_z$ irreps appearing in~(\ref{572su84}), the decomposition of the \mb{572_M}.  Comparing the decompositions given in~(\ref{1682isot}) through (\ref{202isot}) with the decompositions~(\ref{ft1020m2isot}) through (\ref{1a20m2isot}), we see that the situation for baryons in the \mb{(\ohf,\ 572_M)} is somewhat more complicated than for those in the \mb{(\thf,\ 364_S)} because three of the $\su{2}{I}\times\su{4}{T}$ irreps each arise in the decompositions of two of the $SU(8)_{x,y}\times SU(4)_z$ irreps appearing in~(\ref{572su84}).  Both the $\mb{(1,\ 20_M)}_{-1}$ and the $\mb{(0,\ 20_M)}_{-1}$ appear in (\ref{282isot}) and (\ref{362isot}), while the $\mb{(\ohf,\ 20_M)}_{-2}$ appears in (\ref{862isot}) and (\ref{8102isot}).  Without working out the indicial details, one cannot say which linear combinations of these irreps appear in the decompositions~(\ref{ft1020m2isot}), (\ref{ft820m2isot}), and (\ref{1a20m2isot}).  However, the remainder of the analysis goes through as before.  

The $\mb{(\ohf,\ 20_S)}_0$ appears in~(\ref{su842isot}) only once, in the decomposition~(\ref{1682isot}) of the $\mb{(168_M,\ 1)}_0$.  Referring to (\ref{572su84}), we see that all $Z=0$ members of the \mb{572_M} are degenerate with the single-taste nucleon.  Similarly, the appearance in~(\ref{su842isot}) of the $\mb{(0,\ 20_S)}_{-1}$ in the decomposition~(\ref{282isot}) of the $\mb{(28_A,\ 4)}_{-1}$ shows that all members of the $\mb{(28_A,\ 4)}_{-1}$ are degenerate with the single-taste \lam, while the appearance of the $\mb{(1,\ 20_S)}_{-1}$ in the decomposition~(\ref{362isot}) shows that all members of the $\mb{(36_S,\ 4)}_{-1}$ are degenerate with the $\Sigma$.  

Noting the other $\su{2}{I}\times\su{4}{T}$ irreps appearing in the decompositions~(\ref{282isot}) and (\ref{362isot}) and where these irreps appear in the decompositions~(\ref{ft2isot}), we conclude that some linear combination of the $\mb{(1,\ 20_M)}_{-1}$ irreps appearing in the decompositions~(\ref{ft1020m2isot}) and (\ref{ft820m2isot}) of the \mb{(10_S,\ 20_M)} and the \mb{(8_M,\ 20_M)}, some linear combination of the $\mb{(0,\ 20_M)}_{-1}$ irreps appearing in the decompositions~(\ref{ft820m2isot}) and (\ref{1a20m2isot}) of the \mb{(8_M,\ 20_M)} and the \mb{(1_A,\ 20_M)}, and the $\mb{(1,\ \bar 4_A)}_{-1}$ appearing in the decomposition~(\ref{ft84a2isot}) of the \mb{(8_M,\ \bar 4_A)} are degenerate with the $\Lambda$.  In the same way, a second linear combination of the $\mb{(0,\ 20_M)}_{-1}$ irreps appearing in the decompositions~(\ref{ft820m2isot}) and (\ref{1a20m2isot}) of the \mb{(8_M,\ 20_M)} and the \mb{(1_A,\ 20_M)}, a second linear combination of the $\mb{(1,\ 20_M)}_{-1}$ irreps appearing in the decompositions~(\ref{ft1020m2isot}) and (\ref{ft820m2isot}) of the \mb{(10_S,\ 20_M)} and the \mb{(8_M,\ 20_M)}, and the $\mb{(0,\ \bar 4_A)}_{-1}$ appearing in the decomposition~(\ref{ft84a2isot}) of the \mb{(8_M,\ \bar 4_A)} are degenerate with the $\Sigma$.  

For $Z=-2$ baryons in the \mb{572_M}, the situation is very similar to the $Z=-1$ case, with one notable exception:  Although the appearance of $\mb{(\ohf,\ 20_S)}_{-2}$ in~(\ref{ft820s2isot}) and (\ref{8102isot}) shows that members of the $\mb{(8,\ 10_S)}_{-2}$ are degenerate with the single-taste $\Xi$, symmetry does not constrain the mass of the $\mb{(8,\ 6_A)}_{-2}$ to any physical value because no taste-symmetric irrep appears in the decomposition~(\ref{862isot}) of $\mb{(8,\ 6_A)}_{-2}$ under $\su{2}{I}\times\su{4}{T}$.  In our discussion of the fourth case of Table~\ref{cases}, we will find that the mass of this irrep can be obtained by taking an appropriate limit of the mass of the partially quenched $\Lambda$.  For now we simply conclude that some linear combination of the $\mb{(\ohf,\ 20_M)}_{-2}$ irreps appearing in the decompositions~(\ref{ft1020m2isot}) and (\ref{ft820m2isot}) of the \mb{(10_S,\ 20_M)} and the \mb{(8_M,\ 20_M)} is degenerate with the $\Xi$, while the $\mb{(\ohf,\ \bar 4_A)}_{-2}$ appearing in the decompositions~(\ref{ft84a2isot}) and (\ref{862isot}) and a second linear combination of the $\mb{(\ohf,\ 20_M)}_{-2}$ irreps are degenerate but have an unphysical mass.  Below these states are denoted by $\Lambda_s$.  

Finally, the mass of the $Z=-3$ irrep in the decompositions~(\ref{572su84}) and (\ref{202isot}), $\mb{(1,\ 20_M)}=\mb{(0,\ 20_M)_{-3}}$, is not constrained to be physical.  This irrep contains 20 states that are identical to the single-flavor nucleons of the \mb{(10_S,\ 20_M)} except for having strange valence quarks instead of up (or down) valence quarks.  Therefore, the mass of this irrep may be obtained by equating the masses of the valence quarks of the partially quenched nucleon and the mass $m_s$ of the strange quark.  Below this irrep is denoted by $N_s$.  A preliminary estimate based on tree-level chiral perturbation theory suggests that the mass of this state should be roughly 1600 MeV; in progress is work to decrease the large uncertainty in this estimate.  

The first three columns of Table~\ref{continuum} summarize the conclusions reached thus far; for the first two cases listed in Table~\ref{cases}, all the states in the \mb{364_S} have masses that are degenerate with the masses of decuplet baryons, while nearly all the states in the \mb{572_M} have masses that are degenerate with the masses of octet baryons.  The exceptions arise in the second case, for which flavor \su{3}{F} is broken to isospin; in this case one energy level has a mass equal to that of the partially quenched $\Lambda$, while a second energy level has a mass equal to that of the partially quenched nucleon.  The appearance of these resonances reflects the partially quenched nature of staggered QCD at nonzero lattice spacing~\cite{Bernard:2006zw,Sharpe:2006re}.  
\begin{table}
\begin{ruledtabular}
\begin{tabular}{llll}
Case and symmetry for $a=0$ & Irreps for $a=0$ & Expected mass & No. of lattice irreps \\ \hline
$m_x=m_y=m_z=\hat m$ & \mb{(\ohf,\ 572_M)} & $N$ & 16\\
$\su{2}{S}\times\su{12}{f}$ & \mb{(\thf,\ 364_S)} & $\Delta$ & 14\\ \hline
$m_x=m_y=\hat m,\;m_z=m_s$ & \mb{(\ohf,\ 168_M,\ 1)} & $N$ & 12\\
$\su{2}{S}\times\su{8}{x,y}\times\su{4}{z}$ & \mb{(\ohf,\ 28_A,\ 4)} & $\Lambda$ & 12\\
 & \mb{(\ohf,\ 36_S,\ 4)} & $\Sigma$ & 12\\
 & \mb{(\ohf,\ 8,\ 10_S)} & $\Xi$ & 7\\
 & \mb{(\ohf,\ 8,\ 6_A)} & $\Lambda_s$ & 5\\
 & \mb{(\ohf,\ 1,\ 20_M)} & $N_s$ & 4\\
 & \mb{(\thf,\ 120_S,\ 1)} & $\Delta$ & 13\\
 & \mb{(\thf,\ 36_S,\ 4)} & $\Sigma^*$ & 20\\
 & \mb{(\thf,\ 8,\ 10_S)} & $\Xi^*$ & 13\\
 & \mb{(\thf,\ 1,\ 20_S)} & $\Omega^-$ & 7
\end{tabular}
\end{ruledtabular}
\caption{\label{continuum}Expected continuum degeneracies of the irreps of the continuum valence symmetries (first three columns) and the number of lattice irreps appearing in the decomposition of each continuum irrep (last column).}
\end{table}

At nonzero lattice spacing, discretization effects break the continuum symmetry to that of the lattice, lifting degeneracies within the continuum irreps and introducing mixing among states with the same conserved lattice quantum numbers.  Degeneracies and mixings at nonzero lattice spacing are governed by the lattice symmetry.  Decomposing the continuum irreps into irreps of the lattice symmetries reveals the number of splittings that symmetry breaking effects introduce within each continuum irrep and the presence of off-diagonal elements that these discretization effects introduce in the mass matrix.  For the first case of Table~\ref{cases}, the decomposition is most easily obtained by considering
\begin{equation}
\su{2}{S}\times\su{12}{f}\supset\su{2}{S}\times\su{3}{F}\times\su{4}{T}\supset\sugts,\label{c1tolat}
\end{equation} 
while for the second case of Table~\ref{cases}, we consider
\begin{equation}
\su{2}{S}\times\su{8}{x,y}\times\su{4}{z}\supset\su{2}{S}\times\su{2}{I}\times\su{4}{T}\supset\su{2}{I}\times\mr{GTS}.\label{c2tolat}
\end{equation}
Discretization effects break continuum parity-spin-taste $\mc{P}\times\su{2}{S}\times\su{4}{T}$ down to $\mc{P}\times\mr{GTS}$.  The subgroup of \gts\ consisting of cubic rotations is embedded in the diagonal of the direct product of $\su{2}{S}$ and the diagonal $SU(2)$ subgroup of the spinor $SO(4)$ subgroup of $\su{4}{T}$; \gts\ includes cubic rotations, spatial inversion of the lattice, and spatial shifts by one lattice site~\cite{Golterman:1984dn}.  The direct product with continuum parity that occurs in decompositions under the lattice symmetry and the parity quantum numbers of the $\mc{P}\times\mr{GTS}$ irreps are suppressed in (\ref{c1tolat}), (\ref{c2tolat}), and the rest of Sec.'s~\ref{spec} and~\ref{operspec}.  Recalling the relevant decompositions from Table 5 of~\cite{Golterman:1984dn},
\begin{subequations}
\label{taste2gts}
\begin{eqnarray}
SU(2)_S\times SU(4)_T&\supset& \mr{GTS} \nonumber \\
\mb{(\ohf,\ 20_S)}&\rightarrow& \mb{8}\oplus2\mb{(16)}\label{t1}\\
\mb{(\ohf,\ 20_M)}&\rightarrow& 3\mb{(8)\oplus16}\label{t2}\\
\mb{(\ohf,\ \bar 4_A)}&\rightarrow& \mb{8}\label{t3}\\
\mb{(\thf,\ 20_S)}&\rightarrow& 2\mb{(8)}\oplus2\mb{(8^{\prime})}\oplus3\mb{(16)}\label{t4}\\
\mb{(\thf,\ 20_M)}&\rightarrow& \mb{8}\oplus\mb{8^{\prime}}\oplus4\mb{(16)}\label{t5}\\
\mb{(\thf,\ \bar 4_A)}&\rightarrow& \mb{16}.\label{t6}
\end{eqnarray}
\end{subequations}
Combining these results with the decompositions~(\ref{ks2ft}) immediately gives
\begin{subequations}
\label{ks2gts}
\begin{eqnarray}
SU(2)_S\times SU(12)_f&\supset& \sugts \nonumber\\
\mb{(\ohf,\ 572_M)}&\rightarrow& 3\mb{(10_S,\ 8)\oplus(10_S,\ 16)}\oplus5\mb{(8_M,\ 8)}\nonumber\\
&&\oplus\;3\mb{(8_M,\ 16)}\oplus3\mb{(1_A,\ 8)\oplus(1_A,\ 16)}\label{572}\\
\mb{(\thf,\ 364_S)}&\rightarrow& 2\mb{(10_S,\ 8)}\oplus2\mb{(10_S,\ 8^{\prime})}\oplus3\mb{(10_S,\ 16)}\oplus\mb{(8_M,\ 8)}\nonumber\\
&&\oplus\;\mb{(8_M,\ 8^{\prime})}\oplus4\mb{(8_M,\ 16)}\oplus\mb{(1_A,\ 16)}.\label{364}
\end{eqnarray}
\end{subequations}
For degenerate valence quarks, the baryons are degenerate within irreps of \sugts.  A chiral perturbation theory calculation suggests that the splittings between the \sugts\ irreps are about 10-40 MeV, depending on the precise values of the quark masses and lattice spacing~\cite{Bailey:2006ua}.  From~(\ref{572}) and the second and third columns of the first case of Table~\ref{continuum}, we see that there exist 16 nearly degenerate \sugts\ multiplets of staggered baryons with masses that become exactly degenerate in the continuum limit with the nucleon.  Similarly, from~(\ref{364}) and Table~\ref{continuum}, 14 nearly degenerate \sugts\ multiplets of staggered baryons have masses that become exactly degenerate in the continuum limit with the $\Delta$.  These conclusions are summarized by the first two entries of the last column of Table~\ref{continuum}.

The presence of multiple \sugts\ irreps of the same type in the decompositions~(\ref{ks2gts}) implies that the baryon mass matrix will in general contain off-diagonal elements in subspaces corresponding to states with the same conserved \sugts\ quantum numbers, i.e., corresponding states transforming in the same type of \sugts\ irrep.  For example, corresponding members of the \mb{(10_S,\ 8)}'s mix, and the \mb{(10_S,\ 8)} occurs three times in~(\ref{572}) and two times in~(\ref{364}).  Therefore, for each of the 80 members of this \sugts\ irrep, there exists a 5-dimensional submatrix of the mass matrix in which all off-diagonal elements are generically nonzero.  In the continuum limit, the off-diagonal elements vanish and the diagonal elements of three of the five members of each submatrix are degenerate with the nucleon, while two are degenerate with the $\Delta$.  

For the second case of Table~\ref{cases}, the baryons are degenerate within irreps of \isosgts.  Combining the decompositions~(\ref{taste2gts}) with those in~(\ref{su842isot}) gives
\begin{subequations}
\label{su2su842isosgts1}
\begin{eqnarray}
SU(2)_S\;\,\times\;\,SU(8)_{x,y}&\times&SU(4)_z\;\,\supset\;\,SU(2)_I\;\,\times\;\,\mr{GTS}\nonumber\\
\mb{(\ohf,\ 168_M,\ 1)}&\rightarrow&3\mb{(\thf,\ 8)}_0\oplus\mb{(\thf,\ 16)}_0\oplus5\mb{(\ohf,\ 8)}_0\oplus3\mb{(\ohf,\ 16)}_0\label{h1}\\
\mb{(\ohf,\ 28_A,\ 4)}&\rightarrow&4\mb{(1,\ 8)}_{-1}\oplus\mb{(1,\ 16)}_{-1}\oplus4\mb{(0,\ 8)}_{-1}\oplus3\mb{(0,\ 16)}_{-1}\label{h2}\\
\mb{(\ohf,\ 36_S,\ 4)}&\rightarrow&4\mb{(1,\ 8)}_{-1}\oplus3\mb{(1,\ 16)}_{-1}\oplus4\mb{(0,\ 8)}_{-1}\oplus\mb{(0,\ 16)}_{-1}\label{h3}\\
\mb{(\ohf,\ 8,\ 10_S)}&\rightarrow&4\mb{(\ohf,\ 8)}_{-2}\oplus3\mb{(\ohf,\ 16)}_{-2}\label{h4}\\
\mb{(\ohf,\ 8,\ 6_A)}&\rightarrow&4\mb{(\ohf,\ 8)}_{-2}\oplus\mb{(\ohf,\ 16)}_{-2}\label{h5}\\
\mb{(\ohf,\ 1,\ 20_M)}&\rightarrow&3\mb{(0,\ 8)}_{-3}\oplus\mb{(0,\ 16)}_{-3}\label{h6}
\end{eqnarray}
\end{subequations}
for the spin-\ohf\ baryons.  For the spin-\thf\ states we have
\begin{subequations}
\label{su2su842isosgts2}
\begin{eqnarray}
SU(2)_S\;\,\times\;\,SU(8)_{x,y}&\times&SU(4)_z\;\,\supset\;\,SU(2)_I\;\,\times\;\,\mr{GTS} \nonumber\\
\mb{(\thf,\ 120_S,\ 1)}&\rightarrow&2\mb{(\thf,\ 8)}_0\oplus2\mb{(\thf,\ 8^{\prime})}_0\oplus3\mb{(\thf,\ 16)}_0\oplus\mb{(\ohf,\ 8)}_0\nonumber\\
&&\oplus\;\mb{(\ohf,\ 8^{\prime})}_0\oplus4\mb{(\ohf,\ 16)}_0\label{g1}\\
\mb{(\thf,\ 36_S,\ 4)}&\rightarrow&3\mb{(1,\ 8)}_{-1}\oplus3\mb{(1,\ 8^{\prime})}_{-1}\oplus7\mb{(1,\ 16)}_{-1}\oplus\mb{(0,\ 8)}_{-1}\nonumber\\
&&\oplus\;\mb{(0,\ 8^{\prime})}_{-1}\oplus5\mb{(0,\ 16)}_{-1}\label{g2}\\
\mb{(\thf,\ 8,\ 10_S)}&\rightarrow&3\mb{(\ohf,\ 8)}_{-2}\oplus3\mb{(\ohf,\ 8^{\prime})}_{-2}\oplus7\mb{(\ohf,\ 16)}_{-2}\label{g3}\\
\mb{(\thf,\ 1,\ 20_S)}&\rightarrow&2\mb{(0,\ 8)}_{-3}\oplus2\mb{(0,\ 8^{\prime})}_{-3}\oplus3\mb{(0,\ 16)}_{-3}\label{g4}
\end{eqnarray}
\end{subequations}
From the decompositions~(\ref{su2su842isosgts1}) and (\ref{su2su842isosgts2}) and the expected masses of the $SU(2)_S\times SU(8)_{x,y}\times SU(4)_z$ irreps listed in Table~\ref{continuum}, we conclude that there exist 12 nearly degenerate \isosgts\ multiplets of staggered baryons with masses that become exactly degenerate in the continuum limit with the nucleon, 12 that become degenerate with the $\Lambda$, 12 that become degenerate with the $\Sigma$, 7 that become degenerate with the $\Xi$, and so forth.  These conclusions are summarized by the last column of Table~\ref{continuum}.  

Just as for the case of degenerate valence quarks, the presence of multiple \isosgts\ irreps of the same type in~(\ref{su2su842isosgts1}) and (\ref{su2su842isosgts2}) implies that the baryon mass matrix contains off-diagonal elements in subspaces corresponding to states with the same conserved \isosgts\ quantum numbers.  For example, corresponding members of the $\mb{(0,\ 8)}_{-1}$'s mix, and the $\mb{(0,\ 8)}_{-1}$ occurs four times in~(\ref{h2}), four times in~(\ref{h3}), and once in~(\ref{g2}).  So for each of the 8 members of this \isosgts\ irrep, there exists a 9-dimensional submatrix of the mass matrix in which all off-diagonal elements are generically nonzero.  In the continuum limit, the off-diagonal elements vanish and the diagonal elements of four of the nine members of each submatrix are degenerate with the $\Lambda$, four are degenerate with the $\Sigma$, and one, with the $\Sigma^*$.  

We now consider the baryon spectra corresponding to the third and fourth lines of Table~\ref{cases}.  For the rest of this section, all statements about masses apply in the continuum limit.  Begin with the second case in Table~\ref{cases} and imagine increasing $m_x$ and $m_y$ until $m_x=m_y=m_z=m_s$.  This procedure leaves the masses of the $N_s$ and the $\Omega^-$ unchanged, while the restoration of $\su{2}{S}\times\su{12}{f}$ implies the degeneracy of all the members of the \mb{(\ohf,\ 572_M)} and the degeneracy of all members of the \mb{(\thf,\ 364_S)}.  

Now consider decreasing $m_z$ until $m_z=\hat m$.  In this case the valence symmetry is again $SU(8)_{x,y}\times SU(4)_z$, but now we have two valence quarks with the mass of the strange quark and only one valence quark with the mass $\hat m$.  Decreasing $m_z$ changes nothing for the $Z=0$ members of the $\su{2}{S}\times\su{12}{f}$ irreps, so the masses of the $Z=0$ states remain the same:  The $Z=0$ baryons in the \mb{(\ohf,\ 572_M)} remain degenerate with the $N_s$, and those in the \mb{(\thf,\ 364_S)}, with the $\Omega^-$.  However, the masses of the $Z\neq0$ baryons change in a way that is not immediately obvious.  

Consider the $Z=-3$ baryons.  If we vary the masses $m_x$ and $m_y$, the masses of the $Z=-3$ baryons are unchanged.  In particular, decreasing the masses $m_x$ and $m_y$ until $m_x=m_y=m_z=\hat m$, we return to the first case of Table~\ref{cases}, and all members of the \mb{(\ohf,\ 572_M)} are degenerate with the nucleon, while all members of the \mb{(\thf,\ 364_S)} are degenerate with the $\Delta$.  Therefore, for $m_x=m_y=m_s$ and $m_z=\hat m$, the mass of the $Z=-3$ baryon in the \mb{(\ohf,\ 572_M)} is degenerate with the nucleon, while the $Z=-3$ baryon in the \mb{(\thf,\ 364_S)} is degenerate with the $\Delta$.  Taking $m_x=m_y=m_s$ and $m_z=\hat m$ instead of $m_x=m_y=\hat m$ and $m_z=m_s$ switches the masses of the $Z=0$ and $Z=-3$ baryons without changing either the continuum or the lattice valence symmetry groups.  Therefore, the decompositions and related analysis summarized in Table~\ref{continuum} can be immediately applied to the fourth case of Table~\ref{cases}.

Next consider the $Z=-1$ and $Z=-2$ baryons of the \mb{(10_S,\ 20_S)} appearing in the decomposition~(\ref{364ft}) of the \mb{364_S}.  For definiteness, consider one of the single-taste baryons.  For the second case of Table~\ref{cases}, the $Z=-1$ baryons are degenerate with the $\Sigma^*$, while the $Z=-2$ baryons are degenerate with the $\Xi^*$.  The $I_3=1$ member of the isotriplet has quark content $uus$, and the $I_3=\ohf$ member of the isodoublet has quark content $uss$.  From the point of view of the valence quark symmetries, these two states are identical except for the values of the valence quark masses.  Therefore, taking $m_x=m_y=m_s$ and $m_z=\hat m$ instead of $m_x=m_y=\hat m$ and $m_z=m_s$ switches the masses of these two baryons.  The continuum and lattice valence symmetries then imply that changing the quark masses in this way switches the masses of all the $Z=-1$ and $Z=-2$ baryons in the \mb{(\thf,\ 364_S)}.  To summarize, switching the masses of the light and strange valence quarks interchanges the continuum masses of the members of the \mb{(\thf,\ 364_S)}:
\begin{eqnarray}
\Delta&\longleftrightarrow&\Omega^- \label{swap}\\
\Sigma^*&\longleftrightarrow&\Xi^* \nonumber
\end{eqnarray}

Now consider the $Z=-1$ and $Z=-2$ members of the \mb{(\ohf,\ 572_M)}.  First consider the effect of switching the light and strange valence quark masses on the continuum masses of the single-taste baryons in the \mb{(8_M,\ 20_S)}.  The $Z=-1$ isotriplet with $I_3=1$ is degenerate with the $\Sigma$ and has quark content $uus$; the $Z=-2$ isodoublet with $I_3=\ohf$ is degenerate with the $\Xi$ and has quark content $uss$.  As before, these two states differ only in the values of the valence quark masses.  Therefore, we conclude as before that switching the light and strange valence quark masses switches the masses of these states and all states that are degenerate with them.  Finally, consider the $Z=-1$ and $Z=-2$ baryons of the \mb{(8_M,\ \bar 4_A)}.  According to the decompositions~(\ref{282isot}) and (\ref{862isot}) and Table~\ref{continuum}, baryons in the $\mb{(1,\ \bar 4_A)}_{-1}$ are degenerate with the $\Lambda$, and baryons in the $\mb{(\ohf,\ \bar 4_A)}_{-2}$, with the $\Lambda_s$.  Again focusing on a specific taste, consider the $I_3=1$ member of the isotriplet and the $I_3=\ohf$ member of the isodoublet.  The quark content is as before, and again we know that these two states differ only in the values of the valence quark masses.  We conclude that the masses of the $\Lambda$ and the $\Lambda_s$ switch if we switch the values of the light and strange valence quark masses.

In a partially quenched simulation with valence and sea quark masses unrelated, the $\Lambda_s$ and the $\Lambda$ are essentially the same state.  The situation here exactly parallels that for the partially quenched $\Delta$ and the partially quenched $\Omega^-$:  Strictly speaking the valence quarks are different, but the freedom to vary the valence quark masses independently of the sea quark masses means that the masses of the two baryons in each case can be freely switched.  As discussed below in Sec.~\ref{operspec}, this freedom implies that in each case the same operators can be used to extract the masses of both states.

These observations allow us to obtain an estimate of the masses of the $\Lambda_s$ and $N_s$ from chiral perturbation theory:  We have only to calculate, respectively, the masses of the partially quenched $\Lambda$ and nucleon, and then switch the masses of the light and strange valence quarks.  A preliminary tree-level result for $N_s$ was noted above; for the $\Lambda_s$, one finds roughly 1400 MeV.  

In summary, the continuum and lattice valence symmetries are the same for the third and fourth cases as for the first two cases of Table~\ref{cases}, so all the decompositions go through unchanged.  The discussion leading to Table~\ref{continuum} for the first and second cases of Table~\ref{cases} and the accompanying conclusions remain valid if one respectively substitutes the third and fourth cases of Table~\ref{cases} for the first and second and makes the following replacements everywhere:
\begin{eqnarray}
N&\longleftrightarrow&N_s \label{switch}\\
\Lambda&\longleftrightarrow&\Lambda_s \nonumber\\
\Sigma&\longleftrightarrow&\Xi \nonumber\\
\Delta&\longleftrightarrow&\Omega^- \nonumber\\
\Sigma^*&\longleftrightarrow&\Xi^* \nonumber
\end{eqnarray}
Any operator that can be used to extract the mass of one member of the above pairs can also be used to extract the mass of the other.  As shown in Sec.~\ref{operspec}, this observation will prove especially useful for extracting the mass of the $\Xi^*$.

\section{\label{operspec}Symmetry and operator spectra}
An operator transforming irreducibly under the lattice symmetries generically interpolates to all states in the \mb{(\ohf,\ 572_M)}, \mb{(\thf,\ 364_S)}, and higher energy $\su{2}{S}\times\su{12}{f}$ irreps that have the conserved lattice quantum numbers of the operator in question.  The decompositions~(\ref{ks2gts}), (\ref{su2su842isosgts1}), and (\ref{su2su842isosgts2}) reveal the conserved lattice quantum numbers of states in the continuum irreps of Table~\ref{continuum}.  Therefore these decompositions reveal the states in the continuum irreps of Table~\ref{continuum} that are created by operators transforming irreducibly under the lattice symmetries.  

For example, for the first case of Table~\ref{cases}, consider an operator transforming as a member of the \sugts\ irrep \mb{(8_M,\ 8)}.  This irrep appears five times in~(\ref{572}) and once in~(\ref{364}).  According to the discussion of Sec.~\ref{spec} summarized in Table~\ref{continuum}, all members of the \mb{(\ohf,\ 572_M)}, including the members of the five \mb{(8_M,\ 8)}'s appearing in~(\ref{572}), are degenerate, in the continuum limit, with the nucleon, while all members of the \mb{(\thf,\ 364_S)}, including the members of the single \mb{(8_M,\ 8)} of (\ref{364}), are degenerate with the $\Delta$.  At nonzero lattice spacing, spin-taste violations lift the degeneracy of the five \mb{(8_M,\ 8)} irreps and introduce mixing within sets of six states with the same conserved \sugts\ quantum numbers.  The operator in question possesses the conserved \sugts\ quantum numbers of one member of the \mb{(8_M,\ 8)} and interpolates to all six states in the \mb{(\ohf,\ 572_M)} and \mb{(\thf,\ 364_S)} with these quantum numbers; i.e., the operator interpolates to the five corresponding states in the \mb{(8_M,\ 8)}'s appearing in the decomposition~(\ref{572}) of the \mb{(\ohf,\ 572_M)} and the single corresponding state in the \mb{(8_M,\ 8)} appearing in the decomposition~(\ref{364}) of the \mb{(\thf,\ 364_S)}.  

Making analogous observations for each of the lattice irreps appearing in the decompositions~(\ref{ks2gts}), (\ref{su2su842isosgts1}), and (\ref{su2su842isosgts2}) of the continuum irreps of Table~\ref{continuum} gives the number of nondegenerate staggered baryons created by an operator transforming in each lattice irrep that have masses that become degenerate, in the continuum limit, with each of the expected masses listed in Table~\ref{continuum}.  This number is simply equal to the number of times that the lattice irrep of the operator in question occurs in the decomposition of the corresponding continuum irrep.  The results of this analysis are listed in Tables~\ref{os1} and \ref{os2}.  Because a given operator interpolates to states that have the same conserved lattice quantum numbers, states in a given line of Table~\ref{os1} or \ref{os2} are mixed by spin-taste violations.  As for Table~\ref{continuum}, Tables~\ref{os1} and \ref{os2} remain valid if we respectively substitute the quark masses in lines 3 and 4 of Table~\ref{cases} for the quark masses of Tables~\ref{os1} and \ref{os2} and make the substitutions~(\ref{switch}).  
\begin{table}
\begin{ruledtabular}
\begin{tabular}{llc}
Case and symmetry for $a\neq0$ & Operator irreps (for $a\neq0$) & States created/mixed \\ \hline
$m_x=m_y=m_z=\hat m$ & \mb{(10_S,\ 8)} & 3 $N$ and 2 $\Delta$ \\
\sugts\ & \mb{(10_S,\ 8^{\prime})} & 2 $\Delta$ \\
 & \mb{(10_S,\ 16)} & 1 $N$ and 3 $\Delta$ \\
 & \mb{(8_M,\ 8)} & 5 $N$ and 1 $\Delta$ \\
 & \mb{(8_M,\ 8^{\prime})} & 1 $\Delta$ \\
 & \mb{(8_M,\ 16)} & 3 $N$ and 4 $\Delta$ \\
 & \mb{(1_A,\ 8)} & 3 $N$ \\
 & \mb{(1_A,\ 16)} & 1 $N$ and 1 $\Delta$
\end{tabular}
\end{ruledtabular}
\caption{\label{os1}For operators transforming in a given \sugts\ irrep, the number of corresponding nondegenerate baryons in the staggered spectrum that are expected to be degenerate, in the continuum limit, with the listed states.}
\end{table}
\begin{table}
\begin{ruledtabular}
\begin{tabular}{llc}
Case and symmetry for $a\neq0$ & Operator irreps (for $a\neq0$) & States created/mixed \\ \hline
$m_x=m_y=\hat m,\;m_z=m_s$ & $\mb{(\thf,\ 8)}_0$ & 3 $N$ and 2 $\Delta$ \\
\isosgts\ & $\mb{(\thf,\ 8^{\prime})}_0$ & 2 $\Delta$ \\
 & $\mb{(\thf,\ 16)}_0$ & 1 $N$ and 3 $\Delta$ \\
 & $\mb{(\ohf,\ 8)}_0$ & 5 $N$ and 1 $\Delta$ \\
 & $\mb{(\ohf,\ 8^{\prime})}_0$ & 1 $\Delta$ \\
 & $\mb{(\ohf,\ 16)}_0$ & 3 $N$ and 4 $\Delta$ \\
 & $\mb{(1,\ 8)}_{-1}$ & 4 $\Lambda$, 4 $\Sigma$, and 3 $\Sigma^*$ \\
 & $\mb{(1,\ 8^{\prime})}_{-1}$ & 3 $\Sigma^*$ \\
 & $\mb{(1,\ 16)}_{-1}$ & 1 $\Lambda$, 3 $\Sigma$, and 7 $\Sigma^*$ \\
 & $\mb{(0,\ 8)}_{-1}$ & 4 $\Lambda$, 4 $\Sigma$, and 1 $\Sigma^*$ \\
 & $\mb{(0,\ 8^{\prime})}_{-1}$ & 1 $\Sigma^*$ \\
 & $\mb{(0,\ 16)}_{-1}$ & 3 $\Lambda$, 1 $\Sigma$, and 5 $\Sigma^*$ \\
 & $\mb{(\ohf,\ 8)}_{-2}$ & 4 $\Xi$, 4 $\Lambda_s$, and 3 $\Xi^*$ \\
 & $\mb{(\ohf,\ 8^{\prime})}_{-2}$ & 3 $\Xi^*$ \\
 & $\mb{(\ohf,\ 16)}_{-2}$ & 3 $\Xi$, 1 $\Lambda_s$, and 7 $\Xi^*$ \\
 & $\mb{(0,\ 8)}_{-3}$ & 3 $N_s$ and 2 $\Omega^-$ \\
 & $\mb{(0,\ 8^{\prime})}_{-3}$ & 2 $\Omega^-$ \\
 & $\mb{(0,\ 16)}_{-3}$ & 1 $N_s$ and 3 $\Omega^-$
\end{tabular}
\end{ruledtabular}
\caption{\label{os2}For operators transforming in a given \isosgts\ irrep, the number of corresponding nondegenerate baryons in the staggered spectrum that are expected to be degenerate, in the continuum limit, with the listed states.}
\end{table}

Because the splittings and mixings introduced by spin-taste violations in the spectrum of a given operator are probably on the order of a few tens of MeV at currently practical quark masses and lattice spacings, simply ignoring excited states and mixing when fitting to correlators to extract masses may introduce significant systematic errors.  Ideally, one would like to use operators that interpolate to only a few states having very different masses.  Referring to Tables~\ref{os1} and \ref{os2}, two \sugts\ irreps and two \isosgts\ irreps stand out:  the \mb{(8_M,\ 8^{\prime})} and \mb{(1_A,\ 16)} of \sugts\ and the $\mb{(\ohf,\ 8^{\prime})}_0$ and $\mb{(0,\ 8^{\prime})}_{-1}$ of \isosgts.  When used with the quark masses shown in the tables, operators transforming in these irreps interpolate to states that are not part of nearly degenerate lattice multiplets split and mixed by spin-taste violations; therefore, such operators could be used to extract the corresponding baryon masses without needing to account for these splittings and mixings.  

When used with the quark masses of the first line of Table~\ref{cases}, operators transforming in the \mb{(8_M,\ 8^{\prime})} create states in a single $\Delta$ irrep, while operators in the \mb{(1_A,\ 16)} interpolate to a nucleon irrep and another $\Delta$ irrep.  Although operators in the \mb{(1_A,\ 16)} also interpolate to a $\Delta$ irrep, the physical splitting between the nucleon and the $\Delta$ is about 200 MeV, so accounting for this excited state would be much more feasible than accounting for those associated with the spin-taste violations in other channels.  Considering the third case of Table~\ref{cases} and making the substitutions~(\ref{switch}) in Table~\ref{os1}, we see that the \mb{(8_M,\ 8^{\prime})} interpolates to a single $\Omega^-$ irrep, and the \mb{(1_A,\ 16)}, to an $N_s$ irrep and an $\Omega^-$ irrep; as for the nucleon and $\Delta$, channels with splittings and mixings are avoided.  

When flavor \su{3}{F} is broken to isospin, operators transforming in the $\mb{(\ohf,\ 8^{\prime})}_0$ create states in a $\Delta$ irrep, and operators transforming in the $\mb{(0,\ 8^{\prime})}_{-1}$ create states in a $\Sigma^*$ irrep.  Given the results for the case of degenerate valence quarks, the first observation here is not surprising:  The \mb{(8_M,\ 8^{\prime})} appears only once in the decomposition of the \mb{(\thf,\ 364_S)} and, in the decomposition of \sugts\ irreps into \isosgts\ irreps, the $\mb{(\ohf,\ 8^{\prime})}_0$ appears only in the decomposition of the \mb{(8_M,\ 8^{\prime})}.  In contrast, the emergence of the isosinglet $\mb{(0,\ 8^{\prime})}_{-1}$ as a $\Sigma^*$ irrep is counterintuitive; although the isospin is unphysical, one could use an operator in this irrep to cleanly extract the mass of the $\Sigma^*$.  Considering the fourth case of Table~\ref{cases} and making the substitutions~(\ref{switch}) in Table~\ref{os2}, the isosinglet $\mb{(0,\ 8^{\prime})}_{-1}$ interpolates to a single irrep having the mass of the $\Xi^*$.  In summary, by using appropriate operators and quark masses, we can extract the masses of the entire decuplet while completely avoiding the splittings and mixings introduced by spin-taste violations.  

Unfortunately, our ability to vary the valence quark masses and thereby use the same operators to extract the masses of the $\Lambda$ and $\Lambda_s$ is not so useful; the same comment applies to the masses of the $\Sigma$ and $\Xi$ (cf.~(\ref{switch})).  The reason is that there exist no \isosgts\ irreps in Table~\ref{os2} that interpolate to a single $\Sigma$, $\Xi$, $\Lambda$, or $\Lambda_s$.  The complications due to spin-taste violations must be overcome if one is to extract the masses of the $\Sigma$, $\Xi$, and $\Lambda$ with complete control over systematic errors.  

The lattice irreps of operators interpolating to at most one lattice multiplet in each of the \mb{(\ohf,\ 572_M)} and \mb{(\thf,\ 364_S)} irreps are listed again in Table~\ref{best}.  Using operators transforming in these irreps with the valence quark masses shown would allow one to extract the masses of the nucleon, $\Delta$, $\Sigma^*$, $\Xi^*$, and $\Omega^-$ without having to account for splittings and mixings due to spin-taste violations.  These splittings and mixings are completely absent from the spectra of spin-\ohf\ and spin-\thf\ states created by such operators.  
\begin{table}
\begin{ruledtabular}
\begin{tabular}{llc}
Case and symmetry for $a\neq0$ & Operator irreps (for $a\neq0$) & States created/mixed \\ \hline
$m_x=m_y=m_z=\hat m$ & \mb{(8_M,\ 8^{\prime})} & 1 $\Delta$ \\
\sugts\ & \mb{(1_A,\ 16)} & 1 $N$ and 1 $\Delta$ \\ \hline
$m_x=m_y=\hat m,\;m_z=m_s$ & $\mb{(\ohf,\ 8^{\prime})}_0$ & 1 $\Delta$ \\
\isosgts\ & $\mb{(0,\ 8^{\prime})}_{-1}$ & 1 $\Sigma^*$ \\ \hline
$m_x=m_y=m_z=m_s$ & \mb{(8_M,\ 8^{\prime})} & 1 $\Omega^-$ \\
\sugts\ & \mb{(1_A,\ 16)} & 1 $N_s$ and 1 $\Omega^-$ \\ \hline
$m_x=m_y=m_s,\;m_z=\hat m$ & $\mb{(\ohf,\ 8^{\prime})}_0$ & 1 $\Omega^-$ \\
\isosgts\ & $\mb{(0,\ 8^{\prime})}_{-1}$ & 1 $\Xi^*$ \\
\end{tabular}
\end{ruledtabular}
\caption{\label{best}Irreps of operators that interpolate to at most one lattice multiplet in each of the irreps \mb{(\ohf,\ 572_M)} and \mb{(\thf,\ 364_S)}.  The spectra of lightest spin-\ohf\ and spin-\thf\ baryons created by such operators are not split or mixed by spin-taste violations.}
\end{table}

\section{\label{cons}Operators transforming irreducibly under \sugts}

The enumeration parallels that in~\cite{Golterman:1984dn}.  Baryon operators transforming irreducibly under \sugts\ are constructed of three staggered fields, each transforming in the fundamental rep of \sugts.  The baryons are color singlet fermions, so the operators are antisymmetric under permutation of the color indices and under simultaneous permutation of color, flavor, and \mr{GTS} indices.  One way to understand the symmetry requirement on the \mr{GTS} indices is to note that they correspond to the continuum spin and taste indices.  Alternatively, simultaneously interchanging all indices is equivalent to interchanging the staggered fields themselves.

Employing the notation of~\cite{Golterman:1984dn}, one considers the objects
\begin{equation}
_{ijk}{\tilde B}_{ABC}\equiv \sum_{\mb{x},\,x_k\,\mr{even}} \textstyle{\frac{1}{6}}\epsilon_{abc}D_A\chi_i^a(\mb{x})D_B\chi_j^b(\mb{x})D_C\chi_k^c(\mb{x}),\label{B}
\end{equation}
where $i$, $j$, and $k$ are \su{3}{F} indices, $a$, $b$, and $c$ are color indices, $\chi_i^a(\mb{x})$ is a staggered field transforming in the \mb{(3,\ 8)} of \sugts, the sum is over all elementary cubes in the block lattice, $A$, $B$, and $C$ are \mr{GTS} indices, and $D_A$ is a symmetric shift operator defined on the staggered fields by 
\begin{equation}
D_A\chi_i^a(\mb{x})=\ohf[\chi_i^a(\mb{x+a}_A)+\chi_i^a(\mb{x-a}_A)],\label{defnD}
\end{equation}
where $\mb{a}_A$ points to one of the eight corners of an elementary cube.  For each set of \gts\ indices, we apply the reduction rule of~\cite{Golterman:1984dn} to obtain operators that are maximally local:  Whenever two or three symmetric shift operators of the same type act separately on staggered fields in~(\ref{B}), we replace these shift operators with a single shift of the same type acting on the product of the associated staggered fields.  This procedure does not change the transformation properties of the operators under \gts\ or \su{3}{F}.  

The operators ${\tilde B}$ are completely symmetric under simultaneous permutation of flavor and \mr{GTS} indices:  \begin{eqnarray*}
_{ijk}{\tilde B}_{ABC}=\phantom{}_{jki}{\tilde B}_{BCA}=\phantom{}_{kij}{\tilde B}_{CAB}=\phantom{}_{jik}{\tilde B}_{BAC}=\phantom{}_{ikj}{\tilde B}_{ACB}=\phantom{}_{kji}{\tilde B}_{CBA}.
\end{eqnarray*}
Embedding the fundamental rep of \mr{GTS} in the fundamental rep of $SU(8)$, the symmetry of ${\tilde B}$ under simultaneous interchange of flavor and \mr{GTS} indices implies that ${\tilde B}$ transforms in the completely symmetric representation of $SU(24)$, where the fundamental rep of $SU(24)$ coincides with the fundamental rep (the \mb{(3,\ 8)}) of \sugts.  The appearance of the symmetric irrep of $SU(24)$ could have been anticipated from the non-relativistic quark model or the continuum symmetries of the valence sector of staggered QCD; referring to the decompositions~(\ref{symmetric2600}) and (\ref{ks2gts}), the uniqueness of the decompositions implies that operators derived by decomposing the symmetric irrep of $SU(24)$ into irreps of \sugts\ will transform in irreps whose types correspond one-to-one with the \sugts\ irreps of (\ref{ks2gts}).  Hence, the resulting set of operators is complete in the sense that, neglecting contamination from excited states, the operators could be used with matrix fits to extract the masses of all the lightest spin-\ohf\ and spin-\thf\ states.  This completeness also provides a consistency check for the enumeration of the operators:  The decomposition of the symmetric irrep of $SU(24)$ deduced by applying \sugts\ transformations to the independent components of ${\tilde B}$ must match that implied by the decompositions~(\ref{symmetric2600}) and (\ref{ks2gts}).

Decomposing the symmetric rep of $SU(24)$ under $SU(3)_F\times SU(8)$ gives three product irreps:
\begin{eqnarray}
SU(24)&\supset &SU(3)_F\times SU(8)\nonumber\\
\mb{2600_S}&\rightarrow&\mb{(10_S,\ 120_S)\oplus(8_M,\ 168_M)\oplus(1_A,\ 56_A)}\label{24to3x8}
\end{eqnarray}
In terms of the corresponding tensor components, the decomposition~(\ref{24to3x8}) means \[_{ijk}{\tilde B}_{ABC}=\phantom{}_{ijk}\mc{S}_{ABC}+\phantom{}_{ijk}{\mc{M}}_{ABC}+\phantom{}_{ijk}\mc{A}_{ABC},\] where $\mc{S}$ is symmetric under arbitrary, independent permutations of \su{3}{F} and \mr{GTS} indices, $\mc{A}$ is antisymmetric under arbitrary, independent permutations of \su{3}{F} and \mr{GTS} indices, and ${\mc{M}}$ is defined by the relations \begin{eqnarray*}
&_{ijk}{\mc{M}}_{ABC}=\phantom{}_{jki}{\mc{M}}_{BCA}=\phantom{}_{kij}{\mc{M}}_{CAB}=\phantom{}_{jik}{\mc{M}}_{BAC}=\phantom{}_{ikj}{\mc{M}}_{ACB}=\phantom{}_{kji}{\mc{M}}_{CBA}&\\
&_{ijk}{\mc{M}}_{ABC}+\phantom{}_{jki}{\mc{M}}_{ABC}+\phantom{}_{kij}{\mc{M}}_{ABC}=0\phantom{.}&\\
&_{ijk}{\mc{M}}_{ABC}+\phantom{}_{ijk}{\mc{M}}_{BCA}+\phantom{}_{ijk}{\mc{M}}_{CAB}=0.&
\end{eqnarray*}
Hence, $\mc{S}$, $\mc{M}$, and $\mc{A}$ all have the same symmetry as $\tilde B$; they are completely symmetric under simultaneous permutation of \su{3}{F} and \mr{GTS} indices.  By construction, the independent, orthonormal components of $\mc{S}$, $\mc{M}$, and $\mc{A}$ have definite \su{3}{F} quantum numbers.  The reader can verify that\begin{subequations}
\label{2600tosu3xsu8}
\begin{eqnarray}
_{ijk}\mc{S}_{ABC}&=&\osx(_{ijk}{\tilde B}_{ABC}+\phantom{}_{jki}{\tilde B}_{ABC}+\phantom{}_{kij}{\tilde B}_{ABC}+\phantom{}_{jik}{\tilde B}_{ABC}+\phantom{}_{ikj}{\tilde B}_{ABC}+\phantom{}_{kji}{\tilde B}_{ABC})\label{2600toss}\\
_{ijk}{\mc{M}}_{ABC}&=&\otd(2_{ijk}{\tilde B}_{ABC}-\phantom{}_{jki}{\tilde B}_{ABC}-\phantom{}_{kij}{\tilde B}_{ABC})\label{2600tomm}\\
_{ijk}\mc{A}_{ABC}&=&\osx(_{ijk}{\tilde B}_{ABC}+\phantom{}_{jki}{\tilde B}_{ABC}+\phantom{}_{kij}{\tilde B}_{ABC}-\phantom{}_{jik}{\tilde B}_{ABC}-\phantom{}_{ikj}{\tilde B}_{ABC}-\phantom{}_{kji}{\tilde B}_{ABC})\label{2600toaa}.
\end{eqnarray}
\end{subequations}

To proceed further one must decompose these $SU(3)_F\times SU(8)$ irreps into irreps of \sugts.  Following~\cite{Golterman:1984dn}, one identifies a minimal set of independent field components from which all others can be obtained by applying \gts\ transformations.  (See~Eqs.~(5.4) of~\cite{Golterman:1984dn}.)  The action of \gts\ on the underlying staggered fields then dictates the linear combinations of the independent components that transform irreducibly under \sugts.  One can find these irreducible components (irreducibly transforming linear combinations of the independent fields) by considering an arbitrary linear combination of the independent fields, applying the generators of \gts\ to these fields by explicitly transforming the staggered fields, and then fixing the coefficients by demanding that the linear combination transform within one of the irreps of \gts.  Since all group elements can be written as products of the generators, this procedure exhausts the irreducibility constraints on the coefficients.  

In~\cite{Golterman:1984dn}, this analysis was much expedited by two observations:  First, the minimal set of independent components fall into seven distinct geometric classes distinguished by the relative elementary-cube locations where the staggered fields reside.  Because members of different geometric classes are not mixed by the elements of \mr{GTS}, irreducible components must be linear combinations of independent components that belong to the same geometric class.  This observation allows one to restrict the search for irreducible components to subspaces of the 120-dimensional vector space defined by the independent fields; in practice, the largest of these subspaces is only 3-dimensional.  When searching for irreducible components in the 2600-dimensional space considered here, this observation proves its worth many times over; the largest subspaces that one need consider are only 12-dimensional.  

The second observation is that one can uniquely identify irreducible components transforming in the irreps \mb{8}, \mb{8^{\prime}}, and \mb{16} of \mr{GTS} by identifying irreducible components transforming respectively in the irreps $A_1^{\pm}$, $A_2^{\pm}$, and $E^{\pm}$ of the octahedral group $\mr{O_h}\subset\mr{GTS}$; \gts\ is the union of \mr{O_h} and spatial shifts by one lattice site.  Given a set of independent components that is closed under \mr{O_h}, one can ignore the transformation properties under spatial shifts and work exclusively with the generators of \mr{O_h}.  This observation follows from the decomposition of the \gts\ irreps under \mr{O_h}~\cite{Golterman:1984dn}:
\begin{eqnarray*}
\gts&\supset&\mr{O_h}\\
\mb{8\phantom{^{\prime}}}&\rightarrow&A_1^{+}\oplus A_1^{-}\oplus\dots\\
\mb{8^{\prime}}&\rightarrow&A_2^{+}\oplus A_2^{-}\oplus\dots\\
\mb{16}&\rightarrow&E^{+}\oplus E^{-}\oplus\dots,
\end{eqnarray*}
where the neglected irreps of \mr{O_h} do not affect the observation.  Constructing the components ${\tilde B}$ of (\ref{B}) using the symmetric shift operators ${D_A}$ and imposing periodic boundary conditions in the spatial directions, one ensures that the components of ${\tilde B}$ have definite parity under spatial inversion, $I_s$.  Therefore one can find the irreducible components by working with the generators of the cubic rotation group $\mr{O}\subset\mr{O_h}$.  The spatial shifts in \mr{GTS} relate the components of different \mr{O_h} irreps within the same \mr{GTS} irrep; once one has operators transforming irreducibly under \mr{O_h}, applying the shifts to derive the remaining operators in each \gts\ irrep is superfluous because all operators within a given lattice irrep interpolate to precisely the same spectrum.  These observations prove equally useful when constructing operators with \su{3}{F} quantum numbers.

To make these details concrete, we will explicitly search a specific subspace for irreducible components.  First we require the matrix representations of the generators of \mr{O} in each irrep of interest.  The $A$ irreps of \mr{O} are 1-dimensional; in any basis, the matrices of the group elements are equal to the characters.  Choosing $R^{12}$ and $R^{23}$, rotations by $\frac{\pi}{2}$ in the 12- and 23-planes, respectively, as the generators of \mr{O}, and looking up the characters~\cite[p.~127]{Hamermesh}, we have $R^{12}=R^{23}=1$ in the trivial irrep, $A_1$, and $R^{12}=R^{23}=-1$ in $A_2$.  The $E$ irrep is 2-dimensional; to find the matrix representations of $R^{12}$ and $R^{23}$, it suffices to consider an arbitrary (not necessarily irreducible) rep of \mr{O} that contains at least one $E$ in its decomposition into irreps of \mr{O}, project onto the $E$ irrep using the characters~\cite[pp.~111-113]{Hamermesh}, construct the matrices of $R^{12}$ and $R^{23}$, and block diagonalize the results if the projected subspace contains more than one $E$.  Carrying out this procedure and diagonalizing $R^{12}$ gives
\begin{equation}
R^{12}={\begin{pmatrix}
1 &  0 \\
0 & -1
\end{pmatrix}}\quad\quad\text{and}\quad\quad
R^{23}={\begin{pmatrix}
-\ohf & \rthreeotwo \\
\rthreeotwo & \ohf
\end{pmatrix}}.\label{matrix}
\end{equation}

Consider the subspace corresponding to the class 2 operators of~\cite{Golterman:1984dn}; the three staggered fields of such operators reside at two sites within each elementary cube that are separated by one and only one lattice spacing.  Two fields reside at one site, and one, at the other site.  In our case, one set of class 2 operators is given by
\begin{equation}
_{ijk}\mc{S}_{011}\equiv\otd(_{ijk}{\tilde B}_{011}+\phantom{}_{jki}{\tilde B}_{011}+\phantom{}_{kij}{\tilde B}_{011})\label{class2}
\end{equation}
for each $ijk$.  Here we have the \gts\ indices $ABC=011$; with respect to a single elementary cube, one staggered field is evaluated at the origin of the cube, and the other two fields are evaluated at the site removed by one lattice spacing in the positive 1-direction.  For each elementary cube contributing to a component ${\tilde B}$, the symmetric shift operators incorporate fields from neighboring elementary cubes; by definition, the geometric classes are not changed by the symmetric shift operators in ${\tilde B}$~(\ref{B}).  

Applying $I_s$ and the generators of \mr{O} to the staggered fields in the components of ${\tilde B}$ in (\ref{class2}) gives
\begin{subequations}
\label{cubrots}
\begin{eqnarray}
I_s:&\quad&_{ijk}\mc{S}_{011}\rightarrow \phantom{}_{ijk}\mc{S}_{011}\\
R^{12}:&\quad&_{ijk}\mc{S}_{011}\rightarrow \phantom{}_{ijk}\mc{S}_{022}\label{c1}\\
R^{23}:&\quad&_{ijk}\mc{S}_{011}\rightarrow \phantom{}_{ijk}\mc{S}_{011}\label{c4}\\
\nonumber\\
R^{12}:&\quad&_{ijk}\mc{S}_{022}\rightarrow \phantom{}_{ijk}\mc{S}_{011}\label{c2}\\
R^{23}:&\quad&_{ijk}\mc{S}_{022}\rightarrow \phantom{}_{ijk}\mc{S}_{033}\label{c5}\\
\nonumber\\
R^{12}:&\quad&_{ijk}\mc{S}_{033}\rightarrow \phantom{}_{ijk}\mc{S}_{033}\label{c3}\\
R^{23}:&\quad&_{ijk}\mc{S}_{033}\rightarrow \phantom{}_{ijk}\mc{S}_{022}\label{c6}
\end{eqnarray}
\end{subequations}
The sequence of transformations ends because the components $_{ijk}\mc{S}_{011}$, $_{ijk}\mc{S}_{022}$, and $_{ijk}\mc{S}_{033}$ are closed under \mr{O}; these components correspond to a 3-dimensional subspace.  The components $_{ijk}\mc{S}_{022}$ and $_{ijk}\mc{S}_{033}$, like $_{ijk}\mc{S}_{011}$, are invariant under $I_s$ because $\mr{O}\subset\mr{O_h}$; i.e., cubic rotations do not change $I_s$ parity.  

Now we search for irreducible components of \mr{O}.  Consider the linear combination 
\begin{equation}
\alpha\;_{ijk}\mc{S}_{011}+\beta\;_{ijk}\mc{S}_{022}+\gamma\;_{ijk}\mc{S}_{033}.\label{lin}
\end{equation}
We want to find the coefficients $\alpha$, $\beta$, and $\gamma$ such that the resulting linear combination transforms irreducibly under the $A_1$, $A_2$, or $E$ irrep of \mr{O}.  In $A_1$, $R^{12}=R^{23}=1$; components transforming in the $A_1$ irrep are invariant under \mr{O}.  Transforming (\ref{lin}) under $R^{12}$ according to (\ref{c1}), (\ref{c2}), and (\ref{c3}) and demanding invariance gives \[\alpha\;_{ijk}\mc{S}_{022}+\beta\;_{ijk}\mc{S}_{011}+\gamma\;_{ijk}\mc{S}_{033}=\alpha\;_{ijk}\mc{S}_{011}+\beta\;_{ijk}\mc{S}_{022}+\gamma\;_{ijk}\mc{S}_{033},\] so $\alpha=\beta$.  Applying $R^{23}$ in accord with (\ref{c4}), (\ref{c5}), and (\ref{c6}) implies $\beta=\gamma$, and we conclude that \[_{ijk}\mc{S}_{011}+\phantom{}_{ijk}\mc{S}_{022}+\phantom{}_{ijk}\mc{S}_{033}\;\sim\;A_1^+\sim\mb{8};\] the overall normalization does not affect and is therefore not fixed by the transformation properties, and we have a single irreducible component transforming in the $A_1^+$ of \mr{O_h} and therefore the \mb{8} of \gts.  Because we are dealing with a completely reducible 3-dimensional subspace and the \mr{O} irreps have dimensions 1, 2, and 3, we see that the remaining irrep(s) must be either two $A_2$'s or one $E$.  

In $A_2$, $R^{12}=R^{23}=-1$.  Then the transformation of (\ref{lin}) under $R^{12}$ implies that $\alpha=-\beta$ and $\gamma=0$, while that under $R^{23}$ gives $\beta=-\gamma$ and $\alpha=0$; hence we have $\alpha=\beta=\gamma=0$, which means that there exist no irreducible components of $A_2$ in this subspace.  The remaining irrep must be an $E$.  

To find the irreducible components of $E$, we consider the linear combinations
\begin{eqnarray*}
\phi_+&\equiv&\alpha_{+}\>_{ijk}\mc{S}_{011}+\beta_{+}\>_{ijk}\mc{S}_{022}+\gamma_{+}\>_{ijk}\mc{S}_{033}\\
\phi_-&\equiv&\alpha_{-}\>_{ijk}\mc{S}_{011}+\beta_{-}\>_{ijk}\mc{S}_{022}+\gamma_{-}\>_{ijk}\mc{S}_{033}.
\end{eqnarray*}
According to~(\ref{matrix}),
\begin{subequations}
\label{Esearch}
\begin{eqnarray}
R^{12}:&\quad&\phi_+\rightarrow \phi_+\label{E1}\\
R^{23}:&\quad&\phi_+\rightarrow -\ohf\phi_{+}+\rthreeotwo\phi_{-}\label{E3}\\
&\text{and}&\nonumber\\
R^{12}:&\quad&\phi_-\rightarrow -\phi_-\label{E2}\\
R^{23}:&\quad&\phi_-\rightarrow \rthreeotwo\phi_{+}+\ohf\phi_{-}\label{E4}
\end{eqnarray}
\end{subequations}
Transforming $\phi_{+}$ and $\phi_{-}$ under $R^{12}$ according to (\ref{cubrots}) and demanding (\ref{E1}) and (\ref{E2}) gives $\alpha_+=\beta_+$, $\alpha_-=-\beta_-$, and $\gamma_-=0$.  Incorporating this information, transforming $\phi_{+}$ and $\phi_{-}$ under $R^{23}$ according to (\ref{cubrots}), and demanding (\ref{E3}) and (\ref{E4}) then gives $\alpha_-={\sqrt 3}\,\alpha_+$ and $\gamma_+=-2\,\alpha_+$, and we conclude that 
\[_{ijk}\mc{S}_{011}+\phantom{}_{ijk}\mc{S}_{022}-2_{ijk}\mc{S}_{033}\;\sim\;E_1^+\sim\mb{16}\] and \[_{ijk}\mc{S}_{011}-\phantom{}_{ijk}\mc{S}_{022}\;\sim\;E_{-1}^+\sim\mb{16},\] where the subscripts on $E$ indicate the eigenvalue of $R^{12}$.  

We have decomposed the 3-dimensional subspace into irreps of \mr{O_h} and found that for each set of \su{3}{F} indices, the subspace is an $A_1^+\oplus E^+$ of \mr{O_h}; since operators transforming within the same \gts\ irrep create identical spectra and $\mr{O_h}\subset\gts$, we have constructed all the operators with linearly independent spectra in the corresponding $\mb{8}\oplus\mb{16}$ of \gts.  The symmetry of the \su{3}{F} indices then implies that we have found operators transforming in a \mb{(10_S,\ 8)} and a \mb{(10_S,\ 16)} appearing in the decomposition of the \mb{(10_S,\ 120_S)} of $SU(3)_F\times SU(8)$ into irreps of \sugts.  

Comparing these results with those for the class 2 operators of Table 3 of~\cite{Golterman:1984dn}, we see that the irreducible operators obtained from the decomposition of the \mb{(10_S,\ 120_S)} under \sugts\ are a direct generalization of the single-flavor operators given in~\cite{Golterman:1984dn}.  In fact, all the flavor-symmetric operators with linearly independent spectra can be immediately read off from Table 3 of~\cite{Golterman:1984dn}:  One simply adds a flavor index to each staggered field in the single-flavor operator and then symmetrizes the flavor indices to get the result.  This fact may be understood by considering the indicial symmetries involved in the two cases.  In~\cite{Golterman:1984dn}, the objects corresponding to ${\tilde B}$ in (\ref{B}) are completely symmetric in the \mr{GTS} indices.  (See~Eq.~(5.2) of~\cite{Golterman:1984dn}.)  Likewise, the components of $\mc{S}$ are completely symmetric in the \mr{GTS} indices.  The remainder of the analysis of~\cite{Golterman:1984dn} utilized only the transformation properties of the staggered fields under \mr{GTS} and this symmetry under permutation of the \mr{GTS} indices.  Therefore the analysis goes through unchanged for each member of the \mb{10_S} in the \mb{(10_S,\ 120_S)}, and the claimed result follows.  For completeness, the resulting operators are listed in Table~\ref{SYM}, where we have adopted the notation of~\cite{Golterman:1984dn} for the \mr{GTS} indices.  Since pairs of operators transforming in the same $E$ transform in the same \mb{16} of \gts, we have
\begin{eqnarray}
SU(3)_F\times SU(8)&\supset &\sugts\nonumber \\
\mb{(10_S,\ 120_S)}&\rightarrow &5\mb{(10_S,\ 8)}\oplus 2\mb{(10_S,\ 8^{\prime})}\oplus4\mb{(10_S,\ 16)}\label{ssdecomp}
\end{eqnarray}
in accord with Eq.~(5.5) of~\cite{Golterman:1984dn}.  For each member of the \mb{10_S}, the irrep \mb{120_S} of $SU(8)$ corresponds to the 120 operators of~\cite{Golterman:1984dn}; the irrep \mb{120_S} is the reducible rep \mb{[8\times 8\times 8]_S} of \mr{GTS} analyzed in~\cite{Golterman:1984dn}.
\begin{table}
\begin{ruledtabular}
\begin{tabular}{lclc}
Operators & \mr{O_h} irrep & \sugts\ irrep & Class\\ \hline
$_{ijk}\mc{S}_{000}$ & $A_1^{+}$ & \mb{(10_S,\ 8)} & 1\\
$_{ijk}\mc{S}_{011}+\phantom{}_{ijk}\mc{S}_{022}+\phantom{}_{ijk}\mc{S}_{033}$ & $A_1^+$ & \mb{(10_S,\ 8)}&2\\
$_{ijk}\mc{S}_{011}+\phantom{}_{ijk}\mc{S}_{022}-2_{ijk}\mc{S}_{033}$ & $E_1^+$ & \mb{(10_S,\ 16)}&2\\
$_{ijk}\mc{S}_{011}-\phantom{}_{ijk}\mc{S}_{022}$ & $E_{-1}^+$ & \mb{(10_S,\ 16)}&2\\
$_{ijk}\mc{S}_{0,23,23}+\phantom{}_{ijk}\mc{S}_{0,13,13}+\phantom{}_{ijk}\mc{S}_{0,12,12}$ & $A_1^+$ & \mb{(10_S,\ 8)}&3\\
$_{ijk}\mc{S}_{0,23,23}+\phantom{}_{ijk}\mc{S}_{0,13,13}-2_{ijk}\mc{S}_{0,12,12}$ & $E_1^+$ & \mb{(10_S,\ 16)}&3\\
$_{ijk}\mc{S}_{0,23,23}-\phantom{}_{ijk}\mc{S}_{0,13,13}$ & $E_{-1}^+$ & \mb{(10_S,\ 16)}&3\\
$_{ijk}\mc{S}_{1,12,13}-\phantom{}_{ijk}\mc{S}_{2,21,23}+\phantom{}_{ijk}\mc{S}_{3,31,32}$ & $A_2^-$ & \mb{(10_S,\ 8^{\prime})}&4\\
$_{ijk}\mc{S}_{1,12,13}+\phantom{}_{ijk}\mc{S}_{2,21,23}$ & $E_1^-$ & \mb{(10_S,\ 16)}&4\\
$_{ijk}\mc{S}_{1,12,13}-\phantom{}_{ijk}\mc{S}_{2,21,23}-2_{ijk}\mc{S}_{3,31,32}$ & $E_{-1}^-$ & \mb{(10_S,\ 16)}&4\\
$_{ijk}\mc{S}_{0,0,123}$ & $A_{1}^-$ & \mb{(10_S,\ 8)}&5\\
$_{ijk}\mc{S}_{0,1,23}-\phantom{}_{ijk}\mc{S}_{0,2,13}+\phantom{}_{ijk}\mc{S}_{0,3,12}$ & $A_1^-$ & \mb{(10_S,\ 8)}&6\\
$_{ijk}\mc{S}_{0,1,23}-\phantom{}_{ijk}\mc{S}_{0,2,13}-2_{ijk}\mc{S}_{0,3,12}$ & $E_1^-$ & \mb{(10_S,\ 16)}&6\\
$_{ijk}\mc{S}_{0,1,23}+\phantom{}_{ijk}\mc{S}_{0,2,13}$ & $E_{-1}^-$ & \mb{(10_S,\ 16)}&6\\
$_{ijk}\mc{S}_{123}$ & $A_2^-$ & \mb{(10_S,\ 8^{\prime})}&7
\end{tabular}
\end{ruledtabular}
\caption{\label{SYM}Operators transforming within irreps of \sugts\ obtained by decomposing the \mb{(10_S,\ 120_S)} of $SU(3)_F\times SU(8)$.}
\end{table}

The search for irreducible components of \sugts\ in the irrep \mb{(1_A,\ 56_A)} proceeds in almost the same way.  The list of allowed independent components is shorter than the one for \mc{S} because the components of \mc{A} are completely antisymmetric under permutations of \gts\ indices; accordingly, operators from classes 1, 2, 3, and 5 are disallowed (cf.~Table~\ref{SYM}).  Limiting the search to components from classes 4, 6, and 7 and fixing the coefficients in exactly the same manner as before gives the results shown in Table~\ref{aa}.
\begin{table}
\begin{ruledtabular}
\begin{tabular}{lclc}
Operators & \mr{O_h} irrep & \sugts\ irrep & Class\\ \hline
$_{ijk}\mc{A}_{1,12,13}+\phantom{}_{ijk}\mc{A}_{2,21,23}+\phantom{}_{ijk}\mc{A}_{3,31,32}$ & $A_1^-$ & \mb{(1_A,\ 8)}&4\\
$_{ijk}\mc{A}_{1,12,13}+\phantom{}_{ijk}\mc{A}_{2,21,23}-2_{ijk}\mc{A}_{3,31,32}$ & $E_1^-$ & \mb{(1_A,\ 16)}&4\\
$_{ijk}\mc{A}_{1,12,13}-\phantom{}_{ijk}\mc{A}_{2,21,23}$ & $E_{-1}^-$ & \mb{(1_A,\ 16)}&4\\
$_{ijk}\mc{A}_{0,1,23}-\phantom{}_{ijk}\mc{A}_{0,2,13}+\phantom{}_{ijk}\mc{A}_{0,3,12}$ & $A_1^-$ & \mb{(1_A,\ 8)}&6\\
$_{ijk}\mc{A}_{0,1,23}-\phantom{}_{ijk}\mc{A}_{0,2,13}-2_{ijk}\mc{A}_{0,3,12}$ & $E_1^-$ & \mb{(1_A,\ 16)}&6\\
$_{ijk}\mc{A}_{0,1,23}+\phantom{}_{ijk}\mc{A}_{0,2,13}$ & $E_{-1}^-$ & \mb{(1_A,\ 16)}&6\\
$_{ijk}\mc{A}_{123}$ & $A_1^-$ & \mb{(1_A,\ 8)}&7
\end{tabular}
\end{ruledtabular}
\caption{\label{aa}Operators transforming within irreps of \sugts\ obtained by decomposing the \mb{(1_A,\ 56_A)} of $SU(3)_F\times SU(8)$.}
\end{table}
The results of Table~\ref{aa} represent all operators with distinct spectra obtained by decomposing \mb{(1_A,\ 56_A)} into irreps of \sugts; operators obtained from these by applying shifts and rotations transform in the same irreps of \mr{GTS} and therefore add nothing new.  The results of Table~\ref{aa} thus imply that
\begin{eqnarray}
SU(3)_F\times SU(8)&\supset&\sugts \nonumber\\
\mb{(1_A,\,56_A)}&\rightarrow& 3\mb{(1_A,\,8)}\oplus 2\mb{(1_A,\,16)}.\label{aadecomp}
\end{eqnarray}

The search for irreducible operators in the irrep \mb{(8_M,\ 168_M)} proceeds in the same way except for one additional step at the end.  After constructing linear combinations transforming irreducibly under \mr{GTS}, we construct linear combinations that have definite isospin using components transforming within the same type of \mr{GTS} irrep.  This issue did not arise when considering the other $SU(3)_F\times SU(8)$ irreps because the upness, downness, and strangeness suffice to distinguish members within the \mb{10_S} and \mb{1_A} of \su{3}{F}.

The results of the analysis are shown in Tables~\ref{mm2}, \ref{mm3}, \ref{mm5}, \ref{mm7}, \ref{mm4a}, \ref{mm4b}, \ref{mm4c}, \ref{mm6a}, \ref{mm6b}, and~\ref{mm6c}; in all these tables, $i\neq j$ and some care must be taken in interpreting the flavor indices:  For the operators to have the indicated isospin, a flavor index of 3 corresponds to the strange valence quark.  The results in these tables imply that
\begin{eqnarray}
SU(3)_F\times SU(8)&\supset &\sugts\nonumber \\
\mb{(8_M,\ 168_M)}&\rightarrow &6\mb{(8_M,\ 8)}\oplus \mb{(8_M,\ 8^{\prime})}\oplus7\mb{(8_M,\ 16)}.\label{mmdecomp}
\end{eqnarray}
Together with the decompositions (\ref{24to3x8}), (\ref{ssdecomp}), and (\ref{aadecomp}), decomposition (\ref{mmdecomp}) implies that
\begin{eqnarray}
SU(24)&\supset &\sugts \nonumber\\
\mb{2600_S}&\rightarrow&5\mb{(10_S,\ 8)}\oplus 2\mb{(10_S,\ 8^{\prime})}\oplus4\mb{(10_S,\ 16)}\nonumber\\
&&\oplus\;6\mb{(8_M,\ 8)}\oplus \mb{(8_M,\ 8^{\prime})}\oplus7\mb{(8_M,\ 16)}\nonumber\\
&&\oplus\;3\mb{(1_A,\,8)}\oplus 2\mb{(1_A,\,16)}\label{24tosugts}.
\end{eqnarray}
To check this decomposition against our previous results, recall from (\ref{symmetric2600}) that
\begin{eqnarray*}
SU(24)&\supset &\spinks\\
\mb{2600_S}&\rightarrow&\mb{(\thf,\ 364_S)}\oplus\mb{(\ohf,\ 572_M)}.
\end{eqnarray*}
Comparing the decomposition (\ref{24tosugts}) with the direct sum of the decompositions (\ref{572}) and (\ref{364}), we see that the \sugts\ irreps correspond one-to-one, as they must.
\begin{table}
\begin{ruledtabular}
\begin{tabular}{lclc}
Operator(s) & \mr{O_h} irrep & \isosgts\ irrep(s) & $I_3$ \\ \hline
$_{iij}{\mc{M}}_{011}+\phantom{}_{iij}{\mc{M}}_{022}+\phantom{}_{iij}{\mc{M}}_{033}$ & $A_1^{+}$ & $\mb{(\ohf,\ 8)}_{0}$ & $\pm\ohf$ \\
 & & $\mb{(1,\ 8)}_{-1}$ & $\pm 1$ \\
 & & $\mb{(\ohf,\ 8)}_{-2}$ & $\pm\ohf$ \\
$_{312}{\mc{M}}_{011}+\phantom{}_{312}{\mc{M}}_{022}+\phantom{}_{312}{\mc{M}}_{033}$ & $A_1^{+}$ & $\mb{(1,\ 8)}_{-1}$ & $0$ \\
$_{312}{\mc{M}}_{011}+\phantom{}_{312}{\mc{M}}_{022}+\phantom{}_{312}{\mc{M}}_{033}+\dots $ & $A_1^{+}$ & $\mb{(0,\ 8)}_{-1}$ & $0$ \\
$2(_{123}{\mc{M}}_{011}+\phantom{}_{123}{\mc{M}}_{022}+\phantom{}_{123}{\mc{M}}_{033})$ & & & \\ \hline
$_{iij}{\mc{M}}_{011}+\phantom{}_{iij}{\mc{M}}_{022}-2_{iij}{\mc{M}}_{033}$ & $E_1^{+}$ & $\mb{(\ohf,\ 16)}_{0}$ & $\pm\ohf$ \\
 & & $\mb{(1,\ 16)}_{-1}$ & $\pm 1$ \\
 & & $\mb{(\ohf,\ 16)}_{-2}$ & $\pm\ohf$ \\
$_{312}{\mc{M}}_{011}+\phantom{}_{312}{\mc{M}}_{022}-2_{312}{\mc{M}}_{033}$ & $E_1^{+}$ & $\mb{(1,\ 16)}_{-1}$ & $0$ \\
$_{312}{\mc{M}}_{011}+\phantom{}_{312}{\mc{M}}_{022}-2_{312}{\mc{M}}_{033}+\dots $ & $E_1^{+}$ & $\mb{(0,\ 16)}_{-1}$ & $0$ \\
$2(_{123}{\mc{M}}_{011}+\phantom{}_{123}{\mc{M}}_{022}-2_{123}{\mc{M}}_{033})$ & & & \\
$_{iij}{\mc{M}}_{011}-\phantom{}_{iij}{\mc{M}}_{022}$ & $E_{-1}^{+}$ & $\mb{(\ohf,\ 16)}_{0}$ & $\pm\ohf$ \\
 & & $\mb{(1,\ 16)}_{-1}$ & $\pm 1$ \\
 & & $\mb{(\ohf,\ 16)}_{-2}$ & $\pm\ohf$ \\
$_{312}{\mc{M}}_{011}-\phantom{}_{312}{\mc{M}}_{022}$ & $E_{-1}^{+}$ & $\mb{(1,\ 16)}_{-1}$ & $0$ \\
$_{312}{\mc{M}}_{011}-\phantom{}_{312}{\mc{M}}_{022}+2(_{123}{\mc{M}}_{011}-\phantom{}_{123}{\mc{M}}_{022})$ & $E_{-1}^{+}$ & $\mb{(0,\ 16)}_{-1}$ & $0$
\end{tabular}
\end{ruledtabular}
\caption{\label{mm2}Class 2 operators transforming within irreps of \sugts\ obtained by decomposing the \mb{(8_M,\ 168_M)} of $SU(3)_F\times SU(8)$.  The class 2 operators transform within two \sugts\ irreps, an \mb{(8_M,\ 8)} and an \mb{(8_M,\ 16)}.}
\end{table}
\begin{table}
\begin{ruledtabular}
\begin{tabular}{lclc}
Operator(s) & \mr{O_h} irrep & \isosgts\ irrep(s) & $I_3$ \\ \hline
$_{iij}{\mc{M}}_{0,23,23}+\phantom{}_{iij}{\mc{M}}_{0,13,13}+\phantom{}_{iij}{\mc{M}}_{0,12,12}$ & $A_1^{+}$ & $\mb{(\ohf,\ 8)}_{0}$ & $\pm\ohf$ \\
 & & $\mb{(1,\ 8)}_{-1}$ & $\pm 1$ \\
 & & $\mb{(\ohf,\ 8)}_{-2}$ & $\pm\ohf$ \\
$_{312}{\mc{M}}_{0,23,23}+\phantom{}_{312}{\mc{M}}_{0,13,13}+\phantom{}_{312}{\mc{M}}_{0,12,12}$ & $A_1^{+}$ & $\mb{(1,\ 8)}_{-1}$ & $0$ \\
$_{312}{\mc{M}}_{0,23,23}+\phantom{}_{312}{\mc{M}}_{0,13,13}+\phantom{}_{312}{\mc{M}}_{0,12,12}+\dots $ & $A_1^{+}$ & $\mb{(0,\ 8)}_{-1}$ & $0$ \\
$2(_{123}{\mc{M}}_{0,23,23}+\phantom{}_{123}{\mc{M}}_{0,13,13}+\phantom{}_{123}{\mc{M}}_{0,12,12})$  & & &  \\ \hline
$_{iij}{\mc{M}}_{0,23,23}+\phantom{}_{iij}{\mc{M}}_{0,13,13}-2_{iij}{\mc{M}}_{0,12,12}$ & $E_1^{+}$ & $\mb{(\ohf,\ 16)}_{0}$ & $\pm\ohf$ \\
 & & $\mb{(1,\ 16)}_{-1}$ & $\pm 1$ \\
 & & $\mb{(\ohf,\ 16)}_{-2}$ & $\pm\ohf$ \\
$_{312}{\mc{M}}_{0,23,23}+\phantom{}_{312}{\mc{M}}_{0,13,13}-2_{312}{\mc{M}}_{0,12,12}$ & $E_1^{+}$ & $\mb{(1,\ 16)}_{-1}$ & $0$ \\
$_{312}{\mc{M}}_{0,23,23}+\phantom{}_{312}{\mc{M}}_{0,13,13}-2_{312}{\mc{M}}_{0,12,12}+\dots $ & $E_1^{+}$ & $\mb{(0,\ 16)}_{-1}$ & $0$  \\
$2(_{123}{\mc{M}}_{0,23,23}+\phantom{}_{123}{\mc{M}}_{0,13,13}-2_{123}{\mc{M}}_{0,12,12})$ & & & \\
$_{iij}{\mc{M}}_{0,23,23}-\phantom{}_{iij}{\mc{M}}_{0,13,13}$ & $E_{-1}^{+}$ & $\mb{(\ohf,\ 16)}_{0}$ & $\pm\ohf$ \\
 & & $\mb{(1,\ 16)}_{-1}$ & $\pm 1$ \\
 & & $\mb{(\ohf,\ 16)}_{-2}$ & $\pm\ohf$ \\
$_{312}{\mc{M}}_{0,23,23}-\phantom{}_{312}{\mc{M}}_{0,13,13}$ & $E_{-1}^{+}$ & $\mb{(1,\ 16)}_{-1}$ & $0$ \\
$_{312}{\mc{M}}_{0,23,23}-\phantom{}_{312}{\mc{M}}_{0,13,13}+\dots$ & $E_{-1}^{+}$ & $\mb{(0,\ 16)}_{-1}$ & $0$ \\ $2(_{123}{\mc{M}}_{0,23,23}-\phantom{}_{123}{\mc{M}}_{0,13,13})$ & & &
\end{tabular}
\end{ruledtabular}
\caption{\label{mm3}Class 3 operators transforming within irreps of \sugts\ obtained by decomposing the \mb{(8_M,\ 168_M)} of $SU(3)_F\times SU(8)$.  The class 3 operators transform within an \mb{(8_M,\ 8)} and an \mb{(8_M,\ 16)} of \sugts.}
\end{table}
\begin{table}
\begin{ruledtabular}
\begin{tabular}{lclc}
Operator(s) & \mr{O_h} irrep & \isosgts\ irrep(s) & $I_3$ \\ \hline
$_{iij}{\mc{M}}_{123,0,0}$ & $A_1^{-}$ & $\mb{(\ohf,\ 8)}_{0}$ & $\pm\ohf$ \\
 & & $\mb{(1,\ 8)}_{-1}$ & $\pm 1$ \\
 & & $\mb{(\ohf,\ 8)}_{-2}$ & $\pm\ohf$ \\
$_{312}{\mc{M}}_{123,0,0}$ & $A_1^{-}$ & $\mb{(1,\ 8)}_{-1}$ & $0$ \\
$_{312}{\mc{M}}_{123,0,0}+2_{123}{\mc{M}}_{123,0,0}$ & $A_1^{-}$ & $\mb{(0,\ 8)}_{-1}$ & $0$
\end{tabular}
\end{ruledtabular}
\caption{\label{mm5}Class 5 operators transforming within irreps of \sugts\ obtained by decomposing the \mb{(8_M,\ 168_M)} of $SU(3)_F\times SU(8)$.  The class 5 operators transform within a single \sugts\ irrep, an \mb{(8_M,\ 8)}.}
\end{table}
\begin{table}
\begin{ruledtabular}
\begin{tabular}{lclc}
Operator(s) & \mr{O_h} irrep & \isosgts\ irrep(s) & $I_3$ \\ \hline
$_{iij}{\mc{M}}_{123}+2_{iji}{\mc{M}}_{123}$ & $E_{1}^{-}$ & $\mb{(\ohf,\ 16)}_{0}$ & $\pm\ohf$ \\
 & & $\mb{(1,\ 16)}_{-1}$ & $\pm 1$ \\
 & & $\mb{(\ohf,\ 16)}_{-2}$ & $\pm\ohf$ \\
$2_{312}{\mc{M}}_{123}+\phantom{}_{312}{\mc{M}}_{132}+\phantom{}_{312}{\mc{M}}_{312}-\phantom{}_{312}{\mc{M}}_{213}$ & $E_{1}^{-}$ & $\mb{(1,\ 16)}_{-1}$ & $0$ \\
$_{312}{\mc{M}}_{132}+\phantom{}_{312}{\mc{M}}_{312}+\phantom{}_{312}{\mc{M}}_{213}$ & $E_{1}^{-}$ & $\mb{(0,\ 16)}_{-1}$ & $0$ \\
$_{iij}{\mc{M}}_{123}$ & $E_{-1}^{-}$ & $\mb{(\ohf,\ 16)}_{0}$ & $\pm\ohf$ \\
 & & $\mb{(1,\ 16)}_{-1}$ & $\pm 1$ \\
 & & $\mb{(\ohf,\ 16)}_{-2}$ & $\pm\ohf$ \\
$_{312}{\mc{M}}_{123}+\phantom{}_{312}{\mc{M}}_{312}-\phantom{}_{312}{\mc{M}}_{213}$ & $E_{-1}^{-}$ & $\mb{(1,\ 16)}_{-1}$ & $0$ \\
$_{312}{\mc{M}}_{123}-2_{312}{\mc{M}}_{132}-\phantom{}_{312}{\mc{M}}_{312}-\phantom{}_{312}{\mc{M}}_{213}$ & $E_{-1}^{-}$ & $\mb{(0,\ 16)}_{-1}$ & $0$
\end{tabular}
\end{ruledtabular}
\caption{\label{mm7}Class 7 operators transforming within an \mb{(8_M,\ 16)} of \sugts.}
\end{table}
\begin{table}
\begin{ruledtabular}
\begin{tabular}{lclc}
Operator(s) & \mr{O_h} irrep & \isosgts\ irrep(s) & $I_3$ \\ \hline
$_{iij}{\mc{M}}_{1,12,13}+\phantom{}_{iij}{\mc{M}}_{2,21,23}+\phantom{}_{iij}{\mc{M}}_{3,31,32}-\dots$ & $A_1^{-}$ & $\mb{(\ohf,\ 8)}_{0}$ & $\pm\ohf$\\
$(_{iji}{\mc{M}}_{1,12,13}+\phantom{}_{iji}{\mc{M}}_{2,21,23}+\phantom{}_{iji}{\mc{M}}_{3,31,32})$ & & $\mb{(1,\ 8)}_{-1}$ & $\pm 1$ \\
 & & $\mb{(\ohf,\ 8)}_{-2}$ & $\pm\ohf$ \\
$_{312}{\mc{M}}_{1,12,13}+\phantom{}_{312}{\mc{M}}_{2,21,23}+\phantom{}_{312}{\mc{M}}_{3,31,32}-\dots$ & $A_1^{-}$ & $\mb{(1,\ 8)}_{-1}$ & $0$\\
$(_{312}{\mc{M}}_{1,13,12}+\phantom{}_{312}{\mc{M}}_{2,23,21}+\phantom{}_{312}{\mc{M}}_{3,32,31})+\dots$ &  &  & \\
$2[_{312}{\mc{M}}_{13,1,12}+\phantom{}_{312}{\mc{M}}_{23,2,21}+\phantom{}_{312}{\mc{M}}_{32,3,31}-\dots$ &  &  & \\
$(_{312}{\mc{M}}_{12,1,13}+\phantom{}_{312}{\mc{M}}_{21,2,23}+\phantom{}_{312}{\mc{M}}_{31,3,32})]$ &  &  &  \\
$_{312}{\mc{M}}_{1,12,13}+\phantom{}_{312}{\mc{M}}_{2,21,23}+\phantom{}_{312}{\mc{M}}_{3,31,32}-\dots$ & $A_1^{-}$ & $\mb{(0,\ 8)}_{-1}$ & $0$ \\
$(_{312}{\mc{M}}_{1,13,12}+\phantom{}_{312}{\mc{M}}_{2,23,21}+\phantom{}_{312}{\mc{M}}_{3,32,31})$ &  &  &  \\ \hline
$_{iij}{\mc{M}}_{1,12,13}-\phantom{}_{iij}{\mc{M}}_{2,21,23}+\phantom{}_{iij}{\mc{M}}_{3,31,32}+\dots$ & $A_2^{-}$ & $\mb{(\ohf,\ 8^{\prime})}_{0}$ & $\pm\ohf$\\
$(_{iji}{\mc{M}}_{1,12,13}-\phantom{}_{iji}{\mc{M}}_{2,21,23}+\phantom{}_{iji}{\mc{M}}_{3,31,32})$ & & $\mb{(1,\ 8^{\prime})}_{-1}$ & $\pm 1$ \\
 & & $\mb{(\ohf,\ 8^{\prime})}_{-2}$ & $\pm\ohf$ \\
$_{312}{\mc{M}}_{1,12,13}-\phantom{}_{312}{\mc{M}}_{2,21,23}+\phantom{}_{312}{\mc{M}}_{3,31,32}+\dots$ & $A_2^{-}$ & $\mb{(1,\ 8^{\prime})}_{-1}$ & $0$\\
$(_{312}{\mc{M}}_{1,13,12}-\phantom{}_{312}{\mc{M}}_{2,23,21}+\phantom{}_{312}{\mc{M}}_{3,32,31})$ &  &  &  \\
$_{312}{\mc{M}}_{1,12,13}-\phantom{}_{312}{\mc{M}}_{2,21,23}+\phantom{}_{312}{\mc{M}}_{3,31,32}+\dots$ & $A_2^{-}$ & $\mb{(0,\ 8^{\prime})}_{-1}$ & $0$\\
$(_{312}{\mc{M}}_{1,13,12}-\phantom{}_{312}{\mc{M}}_{2,23,21}+\phantom{}_{312}{\mc{M}}_{3,32,31})+\dots$ &  &  & \\
$2[_{312}{\mc{M}}_{13,1,12}-\phantom{}_{312}{\mc{M}}_{23,2,21}+\phantom{}_{312}{\mc{M}}_{32,3,31}+\dots$ &  &  & \\
$(_{312}{\mc{M}}_{12,1,13}-\phantom{}_{312}{\mc{M}}_{21,2,23}+\phantom{}_{312}{\mc{M}}_{31,3,32})]$ &  &  &
\end{tabular}
\end{ruledtabular}
\caption{\label{mm4a}Class 4 operators transforming within an \mb{(8_M,\ 8)} and \mb{(8_M,\ 8^{\prime})} of \sugts.}
\end{table}
\begin{table}
\begin{ruledtabular}
\begin{tabular}{lclc}
Operator(s) & \mr{O_h} irrep & \isosgts\ irrep(s) & $I_3$ \\ \hline
$_{iij}{\mc{M}}_{1,12,13}+\phantom{}_{iij}{\mc{M}}_{2,21,23}-2_{iij}{\mc{M}}_{3,31,32}-\dots$ & $E_1^{-}$ & $\mb{(\ohf,\ 16)}_{0}$ & $\pm\ohf$ \\
$(_{iji}{\mc{M}}_{1,12,13}+\phantom{}_{iji}{\mc{M}}_{2,21,23}-2_{iji}{\mc{M}}_{3,31,32})$ & & $\mb{(1,\ 16)}_{-1}$ & $\pm 1$ \\
 & & $\mb{(\ohf,\ 16)}_{-2}$ & $\pm\ohf$ \\
$_{312}{\mc{M}}_{1,12,13}+\phantom{}_{312}{\mc{M}}_{2,21,23}-2_{312}{\mc{M}}_{3,31,32}-\dots $ & $E_1^{-}$ & $\mb{(1,\ 16)}_{-1}$ & $0$ \\
$(_{312}{\mc{M}}_{1,13,12}+\phantom{}_{312}{\mc{M}}_{2,23,21}-2_{312}{\mc{M}}_{3,32,31})+\dots $ & & & \\
$2[_{312}{\mc{M}}_{13,1,12}+\phantom{}_{312}{\mc{M}}_{23,2,21}-2_{312}{\mc{M}}_{32,3,31}-\dots $ & & & \\
$(_{312}{\mc{M}}_{12,1,13}+\phantom{}_{312}{\mc{M}}_{21,2,23}-2_{312}{\mc{M}}_{31,3,32})]$ & & & \\
$_{312}{\mc{M}}_{1,12,13}+\phantom{}_{312}{\mc{M}}_{2,21,23}-2_{312}{\mc{M}}_{3,31,32}-\dots $ & $E_1^{-}$ & $\mb{(0,\ 16)}_{-1}$ & $0$ \\
$(_{312}{\mc{M}}_{1,13,12}+\phantom{}_{312}{\mc{M}}_{2,23,21}-2_{312}{\mc{M}}_{3,32,31}) $ & & & \\
$_{iij}{\mc{M}}_{1,12,13}+\phantom{}_{iij}{\mc{M}}_{2,21,23}+\dots$& $E_1^{-}$ & $\mb{(\ohf,\ 16)}_{0}$ & $\pm\ohf$ \\
$(_{iji}{\mc{M}}_{1,12,13}+\phantom{}_{iji}{\mc{M}}_{2,21,23})$ & & $\mb{(1,\ 16)}_{-1}$ & $\pm 1$ \\
 & & $\mb{(\ohf,\ 16)}_{-2}$ & $\pm\ohf$ \\
$_{312}{\mc{M}}_{1,12,13}+\phantom{}_{312}{\mc{M}}_{2,21,23}+\dots$& $E_1^{-}$ & $\mb{(1,\ 16)}_{-1}$ & $0$ \\
$(_{312}{\mc{M}}_{1,13,12}+\phantom{}_{312}{\mc{M}}_{2,23,21}) $&&&\\
$_{312}{\mc{M}}_{1,12,13}+\phantom{}_{312}{\mc{M}}_{2,21,23}+\dots$& $E_1^{-}$ & $\mb{(0,\ 16)}_{-1}$ & $0$ \\
$(_{312}{\mc{M}}_{1,13,12}+\phantom{}_{312}{\mc{M}}_{2,23,21})+\dots $ & & & \\
$2[_{312}{\mc{M}}_{13,1,12}+\phantom{}_{312}{\mc{M}}_{23,2,21}+\dots$& & & \\
$(_{312}{\mc{M}}_{12,1,13}+\phantom{}_{312}{\mc{M}}_{21,2,23})]$ & & & \\
\end{tabular}
\end{ruledtabular}
\caption{\label{mm4b}Class 4, $E_1^{-}$ operators transforming within two \mb{(8_M,\ 16)}'s of \sugts.}
\end{table}
\begin{table}
\begin{ruledtabular}
\begin{tabular}{lclc}
Operator(s) & \mr{O_h} irrep & \isosgts\ irrep(s) & $I_3$ \\ \hline
$_{iij}{\mc{M}}_{1,12,13}-\phantom{}_{iij}{\mc{M}}_{2,21,23}-\dots$ & $E_{-1}^{-}$ & $\mb{(\ohf,\ 16)}_{0}$ & $\pm\ohf$ \\
$(_{iji}{\mc{M}}_{1,12,13}-\phantom{}_{iji}{\mc{M}}_{2,21,23})$ & & $\mb{(1,\ 16)}_{-1}$ & $\pm 1$ \\
 & & $\mb{(\ohf,\ 16)}_{-2}$ & $\pm\ohf$ \\
$_{312}{\mc{M}}_{1,12,13}-\phantom{}_{312}{\mc{M}}_{2,21,23}-\dots $ & $E_{-1}^{-}$ & $\mb{(1,\ 16)}_{-1}$ & $0$ \\
$(_{312}{\mc{M}}_{1,13,12}-\phantom{}_{312}{\mc{M}}_{2,23,21})+\dots $ & & & \\
$2[_{312}{\mc{M}}_{13,1,12}-\phantom{}_{312}{\mc{M}}_{23,2,21}-\dots $ & & & \\
$(_{312}{\mc{M}}_{12,1,13}-\phantom{}_{312}{\mc{M}}_{21,2,23})]$ & & & \\
$_{312}{\mc{M}}_{1,12,13}-\phantom{}_{312}{\mc{M}}_{2,21,23}-\dots $ & $E_{-1}^{-}$ & $\mb{(0,\ 16)}_{-1}$ & $0$ \\
$(_{312}{\mc{M}}_{1,13,12}-\phantom{}_{312}{\mc{M}}_{2,23,21}) $ & & & \\
$_{iij}{\mc{M}}_{1,12,13}-\phantom{}_{iij}{\mc{M}}_{2,21,23}-2_{iij}{\mc{M}}_{3,31,32}+\dots$& $E_{-1}^{-}$ & $\mb{(\ohf,\ 16)}_{0}$ & $\pm\ohf$ \\
$(_{iji}{\mc{M}}_{1,12,13}-\phantom{}_{iji}{\mc{M}}_{2,21,23}-2_{iji}{\mc{M}}_{3,31,32})$ & & $\mb{(1,\ 16)}_{-1}$ & $\pm 1$ \\
 & & $\mb{(\ohf,\ 16)}_{-2}$ & $\pm\ohf$ \\
$_{312}{\mc{M}}_{1,12,13}-\phantom{}_{312}{\mc{M}}_{2,21,23}-2_{312}{\mc{M}}_{3,31,32}+\dots$& $E_{-1}^{-}$ & $\mb{(1,\ 16)}_{-1}$ & $0$ \\
$(_{312}{\mc{M}}_{1,13,12}-\phantom{}_{312}{\mc{M}}_{2,23,21}-2_{312}{\mc{M}}_{3,32,31})$&&&\\
$_{312}{\mc{M}}_{1,12,13}-\phantom{}_{312}{\mc{M}}_{2,21,23}-2_{312}{\mc{M}}_{3,31,32}+\dots$& $E_{-1}^{-}$ & $\mb{(0,\ 16)}_{-1}$ & $0$ \\
$(_{312}{\mc{M}}_{1,13,12}-\phantom{}_{312}{\mc{M}}_{2,23,21}-2_{312}{\mc{M}}_{3,32,31})+\dots$&&&\\
$2[_{312}{\mc{M}}_{13,1,12}-\phantom{}_{312}{\mc{M}}_{23,2,21}-2_{312}{\mc{M}}_{32,3,31}+\dots$& & & \\
$(_{312}{\mc{M}}_{12,1,13}-\phantom{}_{312}{\mc{M}}_{21,2,23}-2_{312}{\mc{M}}_{31,3,32})]$ & & & \\
\end{tabular}
\end{ruledtabular}
\caption{\label{mm4c}Class 4, $E_{-1}^{-}$ operators transforming within two \mb{(8_M,\ 16)}'s of \sugts.  Each operator transforms in the same \mb{(8_M,\ 16)} as the operator on the corresponding lines of Table~\ref{mm4b}.}
\end{table}
\begin{table}
\begin{ruledtabular}
\begin{tabular}{lclc}
Operator(s) & \mr{O_h} irrep & \isosgts\ irrep(s) & $I_3$ \\ \hline
$_{iij}{\mc{M}}_{0,1,23}-\phantom{}_{iij}{\mc{M}}_{0,2,13}+\phantom{}_{iij}{\mc{M}}_{0,3,12}$ & $A_1^{-}$ & $\mb{(\ohf,\ 8)}_{0}$ & $\pm\ohf$ \\
 & & $\mb{(1,\ 8)}_{-1}$ & $\pm 1$ \\
 & & $\mb{(\ohf,\ 8)}_{-2}$ & $\pm\ohf$ \\
$_{312}{\mc{M}}_{0,1,23}-\phantom{}_{312}{\mc{M}}_{0,2,13}+\phantom{}_{312}{\mc{M}}_{0,3,12}+\dots$ & $A_1^{-}$ & $\mb{(1,\ 8)}_{-1}$ & $0$ \\
$(_{312}{\mc{M}}_{23,0,1}-\phantom{}_{312}{\mc{M}}_{13,0,2}+\phantom{}_{312}{\mc{M}}_{12,0,3})-\dots$ & & & \\
$(_{312}{\mc{M}}_{1,0,23}-\phantom{}_{312}{\mc{M}}_{2,0,13}+\phantom{}_{312}{\mc{M}}_{3,0,12})$ & & & \\
$_{312}{\mc{M}}_{0,1,23}-\phantom{}_{312}{\mc{M}}_{0,2,13}+\phantom{}_{312}{\mc{M}}_{0,3,12}+\dots$ & $A_1^{-}$ & $\mb{(0,\ 8)}_{-1}$ & $0$ \\
$(_{312}{\mc{M}}_{23,0,1}-\phantom{}_{312}{\mc{M}}_{13,0,2}+\phantom{}_{312}{\mc{M}}_{12,0,3})+\dots$ & & & \\
$(_{312}{\mc{M}}_{1,0,23}-\phantom{}_{312}{\mc{M}}_{2,0,13}+\phantom{}_{312}{\mc{M}}_{3,0,12})$ & & & \\
$_{iji}{\mc{M}}_{0,1,23}-\phantom{}_{iji}{\mc{M}}_{0,2,13}+\phantom{}_{iji}{\mc{M}}_{0,3,12}$ & $A_1^{-}$ & $\mb{(\ohf,\ 8)}_{0}$ & $\pm\ohf$ \\
 & & $\mb{(1,\ 8)}_{-1}$ & $\pm 1$ \\
 & & $\mb{(\ohf,\ 8)}_{-2}$ & $\pm\ohf$ \\
$_{312}{\mc{M}}_{0,23,1}-\phantom{}_{312}{\mc{M}}_{0,13,2}+\phantom{}_{312}{\mc{M}}_{0,12,3}-\dots$ & $A_1^{-}$ & $\mb{(1,\ 8)}_{-1}$ & $0$ \\
$(_{312}{\mc{M}}_{23,0,1}-\phantom{}_{312}{\mc{M}}_{13,0,2}+\phantom{}_{312}{\mc{M}}_{12,0,3})+\dots$ & & & \\
$(_{312}{\mc{M}}_{1,0,23}-\phantom{}_{312}{\mc{M}}_{2,0,13}+\phantom{}_{312}{\mc{M}}_{3,0,12})$ & & & \\
$_{312}{\mc{M}}_{0,23,1}-\phantom{}_{312}{\mc{M}}_{0,13,2}+\phantom{}_{312}{\mc{M}}_{0,12,3}+\dots$ & $A_1^{-}$ & $\mb{(0,\ 8)}_{-1}$ & $0$ \\
$(_{312}{\mc{M}}_{23,0,1}-\phantom{}_{312}{\mc{M}}_{13,0,2}+\phantom{}_{312}{\mc{M}}_{12,0,3})+\dots$ & & & \\
$(_{312}{\mc{M}}_{1,0,23}-\phantom{}_{312}{\mc{M}}_{2,0,13}+\phantom{}_{312}{\mc{M}}_{3,0,12})$ & & &
\end{tabular}
\end{ruledtabular}
\caption{\label{mm6a}Class 6 operators transforming within two \mb{(8_M,\ 8)}'s of \sugts.}
\end{table}
\begin{table}
\begin{ruledtabular}
\begin{tabular}{lclc}
Operator(s) & \mr{O_h} irrep & \isosgts\ irrep(s) & $I_3$ \\ \hline
$_{iij}{\mc{M}}_{0,1,23}-\phantom{}_{iij}{\mc{M}}_{0,2,13}-2_{iij}{\mc{M}}_{0,3,12}$ & $E_1^{-}$ & $\mb{(\ohf,\ 16)}_{0}$ & $\pm\ohf$ \\
 & & $\mb{(1,\ 16)}_{-1}$ & $\pm 1$ \\
 & & $\mb{(\ohf,\ 16)}_{-2}$ & $\pm\ohf$ \\
$_{312}{\mc{M}}_{0,1,23}-\phantom{}_{312}{\mc{M}}_{0,2,13}-2_{312}{\mc{M}}_{0,3,12}+\dots$ & $E_1^{-}$ & $\mb{(1,\ 16)}_{-1}$ & $0$ \\
$(_{312}{\mc{M}}_{23,0,1}-\phantom{}_{312}{\mc{M}}_{13,0,2}-2_{312}{\mc{M}}_{12,0,3})-\dots$ & & & \\
$(_{312}{\mc{M}}_{1,0,23}-\phantom{}_{312}{\mc{M}}_{2,0,13}-2_{312}{\mc{M}}_{3,0,12})$ & & & \\
$_{312}{\mc{M}}_{0,1,23}-\phantom{}_{312}{\mc{M}}_{0,2,13}-2_{312}{\mc{M}}_{0,3,12}+\dots$ & $E_1^{-}$ & $\mb{(0,\ 16)}_{-1}$ & $0$ \\
$(_{312}{\mc{M}}_{23,0,1}-\phantom{}_{312}{\mc{M}}_{13,0,2}-2_{312}{\mc{M}}_{12,0,3})+\dots$ & & & \\
$(_{312}{\mc{M}}_{1,0,23}-\phantom{}_{312}{\mc{M}}_{2,0,13}-2_{312}{\mc{M}}_{3,0,12})$ & & & \\
$_{iji}{\mc{M}}_{0,1,23}-\phantom{}_{iji}{\mc{M}}_{0,2,13}-2_{iji}{\mc{M}}_{0,3,12}$ & $E_1^{-}$ & $\mb{(\ohf,\ 16)}_{0}$ & $\pm\ohf$ \\
 & & $\mb{(1,\ 16)}_{-1}$ & $\pm 1$ \\
 & & $\mb{(\ohf,\ 16)}_{-2}$ & $\pm\ohf$ \\
$_{312}{\mc{M}}_{0,23,1}-\phantom{}_{312}{\mc{M}}_{0,13,2}-2_{312}{\mc{M}}_{0,12,3}-\dots$ & $E_1^{-}$ & $\mb{(1,\ 16)}_{-1}$ & $0$ \\
$(_{312}{\mc{M}}_{23,0,1}-\phantom{}_{312}{\mc{M}}_{13,0,2}-2_{312}{\mc{M}}_{12,0,3})+\dots$ & & & \\
$(_{312}{\mc{M}}_{1,0,23}-\phantom{}_{312}{\mc{M}}_{2,0,13}-2_{312}{\mc{M}}_{3,0,12})$ & & & \\
$_{312}{\mc{M}}_{0,23,1}-\phantom{}_{312}{\mc{M}}_{0,13,2}-2_{312}{\mc{M}}_{0,12,3}+\dots$ & $E_1^{-}$ & $\mb{(0,\ 16)}_{-1}$ & $0$ \\
$(_{312}{\mc{M}}_{23,0,1}-\phantom{}_{312}{\mc{M}}_{13,0,2}-2_{312}{\mc{M}}_{12,0,3})+\dots$ & & & \\
$(_{312}{\mc{M}}_{1,0,23}-\phantom{}_{312}{\mc{M}}_{2,0,13}-2_{312}{\mc{M}}_{3,0,12})$ & & & \\
\end{tabular}
\end{ruledtabular}
\caption{\label{mm6b}Class 6, $E_1^{-}$ operators transforming within two \mb{(8_M,\ 16)}'s of \sugts.}
\end{table}
\begin{table}
\begin{ruledtabular}
\begin{tabular}{lclc}
Operator(s) & \mr{O_h} irrep & \isosgts\ irrep(s) & $I_3$ \\ \hline
$_{iij}{\mc{M}}_{0,1,23}+\phantom{}_{iij}{\mc{M}}_{0,2,13}$ & $E_{-1}^{-}$ & $\mb{(\ohf,\ 16)}_{0}$ & $\pm\ohf$ \\
 & & $\mb{(1,\ 16)}_{-1}$ & $\pm 1$ \\
 & & $\mb{(\ohf,\ 16)}_{-2}$ & $\pm\ohf$ \\
$_{312}{\mc{M}}_{0,1,23}+\phantom{}_{312}{\mc{M}}_{0,2,13}+\dots$ & $E_{-1}^{-}$ & $\mb{(1,\ 16)}_{-1}$ & $0$ \\
$(_{312}{\mc{M}}_{23,0,1}+\phantom{}_{312}{\mc{M}}_{13,0,2})-\dots$ & & & \\
$(_{312}{\mc{M}}_{1,0,23}+\phantom{}_{312}{\mc{M}}_{2,0,13})$ & & & \\
$_{312}{\mc{M}}_{0,1,23}+\phantom{}_{312}{\mc{M}}_{0,2,13}+\dots$ & $E_{-1}^{-}$ & $\mb{(0,\ 16)}_{-1}$ & $0$ \\
$(_{312}{\mc{M}}_{23,0,1}+\phantom{}_{312}{\mc{M}}_{13,0,2})+\dots$ & & & \\
$(_{312}{\mc{M}}_{1,0,23}+\phantom{}_{312}{\mc{M}}_{2,0,13})$ & & & \\
$_{iji}{\mc{M}}_{0,1,23}+\phantom{}_{iji}{\mc{M}}_{0,2,13}$ & $E_{-1}^{-}$ & $\mb{(\ohf,\ 16)}_{0}$ & $\pm\ohf$ \\
 & & $\mb{(1,\ 16)}_{-1}$ & $\pm 1$ \\
 & & $\mb{(\ohf,\ 16)}_{-2}$ & $\pm\ohf$ \\
$_{312}{\mc{M}}_{0,23,1}+\phantom{}_{312}{\mc{M}}_{0,13,2}-\dots$ & $E_{-1}^{-}$ & $\mb{(1,\ 16)}_{-1}$ & $0$ \\
$(_{312}{\mc{M}}_{23,0,1}+\phantom{}_{312}{\mc{M}}_{13,0,2})+\dots$ & & & \\
$(_{312}{\mc{M}}_{1,0,23}+\phantom{}_{312}{\mc{M}}_{2,0,13})$ & & & \\
$_{312}{\mc{M}}_{0,23,1}+\phantom{}_{312}{\mc{M}}_{0,13,2}+\dots$ & $E_{-1}^{-}$ & $\mb{(0,\ 16)}_{-1}$ & $0$ \\
$(_{312}{\mc{M}}_{23,0,1}+\phantom{}_{312}{\mc{M}}_{13,0,2})+\dots$ & & & \\
$(_{312}{\mc{M}}_{1,0,23}+\phantom{}_{312}{\mc{M}}_{2,0,13})$ & & & \\
\end{tabular}
\end{ruledtabular}
\caption{\label{mm6c}Class 6, $E_{-1}^{-}$ operators transforming within two \mb{(8_M,\ 16)}'s of \sugts.  Each operator transforms in the same \mb{(8_M,\ 16)} as the operator on the corresponding lines of Table~\ref{mm6b}.}
\end{table}

As an example in the use of the tables, consider the third line of Table~\ref{aa}.  The given operator transforms within the \mb{(1_A,\ 16)} of \sugts\ and therefore, according to the discussion of Sec.~\ref{operspec} summarized in Table~\ref{best}, could be used to extract the masses of the nucleon, $\Delta$, $N_s$, and $\Omega^-$ without needing to account for splittings and mixings due to spin-taste violations.  Noting that the operator is a flavor singlet, we consider \[_{123}\mc{A}_{1,12,13}-\phantom{}_{123}\mc{A}_{2,21,23}=\osx\epsilon^{ijk}(_{ijk}{\tilde B}_{1,12,13}-\phantom{}_{ijk}{\tilde B}_{2,21,23}).\]
Unpacking the components of ${\tilde B}$ using (\ref{B}) and applying the reduction rule gives
\begin{eqnarray*}
\sum _{\mathbf{x},\,x_k\;\mathrm{even}}\textstyle{\frac{1}{6}}\epsilon^{ijk} \textstyle{\frac{1}{6}}\epsilon_{abc}&\Biggl\{\sum _{\varepsilon_2\varepsilon_3} \chi_i^a(\mathbf{x}+\mathbf{a}_1)\chi_j^b(\mathbf{x+a}_1+\varepsilon_2\mathbf{a}_2)\chi_k^c(\mathbf{x+a}_1+\varepsilon_3\mathbf{a}_3)&\\
&-\sum _{\varepsilon_1\varepsilon_3}\chi_i^a(\mathbf{x+a}_2)\chi_j^b(\mathbf{x}+\varepsilon_1\mathbf{a}_1+\mathbf{a}_2)\chi_k^c(\mathbf{x+a}_2+\varepsilon_3\mathbf{a}_3)\Biggr\}&
\end{eqnarray*}
where the symmetric shifts have been taken into account by summing over $\varepsilon_{1,2,3}=\pm1$.  

Obtaining the gauge invariant counterpart of this operator is straightforward.  Adding gauge links to the shortest paths in the elementary cube that connect the staggered fields~\cite{Golterman:1984dn} gives
\begin{eqnarray*}
\sum _{\mathbf{x},\,x_k\;\mathrm{even}}\textstyle{\frac{1}{6}}\epsilon^{ijk} \textstyle{\frac{1}{6}}\epsilon_{abc}&\Biggl\{\sum _{\varepsilon_2\varepsilon_3} \chi_i^a(\mathbf{x}+\mathbf{a}_1)[U(\mathbf{x}+\mathbf{a}_1,\,\mathbf{x}+\mathbf{a}_1+\varepsilon_2\mathbf{a}_2)\chi_j(\mathbf{x+a}_1+\varepsilon_2\mathbf{a}_2)]^b\times&\\
&[U(\mathbf{x+a}_1,\,\mathbf{x+a}_1+\varepsilon_3\mathbf{a}_3)\chi_k(\mathbf{x+a}_1+\varepsilon_3\mathbf{a}_3)]^c&\\
&-\sum _{\varepsilon_1\varepsilon_3}\chi_i^a(\mathbf{x+a}_2)[U(\mathbf{x+a}_2,\,\mathbf{x}+\varepsilon_1\mathbf{a}_1+\mathbf{a}_2)\chi_j(\mathbf{x}+\varepsilon_1\mathbf{a}_1+\mathbf{a}_2)]^b\times&\\
&[U(\mathbf{x+a}_2,\,\mathbf{x+a}_2+\varepsilon_3\mathbf{a}_3)\chi_k(\mathbf{x+a}_2+\varepsilon_3\mathbf{a}_3)]^c\Biggr\}&
\end{eqnarray*}
Gauge invariant operators transforming in the other lattice irreps of Table~\ref{best} can be obtained similarly.  The relevant results are listed in Tables~\ref{aa} and \ref{mm4a}.  Although trivial for the above example, averaging gauge links over all shortest paths can be used to maintain cubic covariance.  

\section{\label{conclu}Conclusions}
For the cases given in Table~\ref{cases}, we have examined the utility and limitations of operators transforming in irreps of \sugts\ and identified the continuum limits of the masses of the states created by such operators.  A set of operators transforming irreducibly under \sugts\ was constructed by decomposing the \mb{2600_S} irrep of $SU(24)$ under \sugts.  The operators of this set are linearly independent and transform within irreps that correspond one-to-one with the types of irreps contained in the decompositions of the continuum spin-flavor irreps of the lightest spin-\ohf\ and spin-\thf\ staggered baryons.  Therefore, these operators could be used with matrix fits to extract the masses of all these states.  In practice, the splittings and mixings in the spectrum could render this program difficult.  

However, for the nucleon, $\Delta$, $\Sigma^*$, $\Xi^*$, and $\Omega^-$, there exist operators that can be used to interpolate to isolated staggered states; the splittings and mixings introduced by spin-taste breaking discretization effects are thus avoided, and one should be able to cleanly extract the masses of these baryons.  The required quark masses and irreps are given in Table~\ref{best}, while operators transforming in these irreps are listed in Tables~\ref{aa} and \ref{mm4a}.  We are planning to test these operators for the extraction of physics in the near future~\cite{Bailey:coming}.  

It turns out that all the operators transforming in the irreps of Table~\ref{best} possess unphysical flavor structure.  The interpolating field for the nucleon is antisymmetric in flavor, that for the $\Delta$ and $\Omega^-$ transforms within an $SU(3)_F$ octet, and that for the $\Sigma^*$ and $\Xi^*$ is an isosinglet.  This counterintuitive state of affairs is possible because, in the continuum limit, we can use taste degrees of freedom instead of flavor degrees of freedom to construct physical states.  At nonzero lattice spacing, the flavor irreps must be combined with the \mr{GTS} irreps so that the combined product irreps have the spin-flavor symmetry of the valence sector of 2+1 flavor staggered QCD, a valence sector containing four quark tastes for each of the three quark flavors.  Although the most numerically convenient baryons have unphysical flavor structure, the conclusion that they become degenerate in the continuum limit with the physical nucleon and decuplet baryons rests upon the existence of a transparently physical subspace of baryons, which have physical flavor structure by definition.  The need to identify such a subspace of baryons is a manifestation of the ``valence rooting issue''~\cite{Sharpe:2006re}; the existence of this subspace rests upon the assumption that taste symmetry is restored in the continuum limit, in which case the valence quark tastes become equivalent to physical quark flavors.  Taste restoration is in turn supported by increasingly cogent analytic~\cite{Shamir:2006nj} and numerical evidence~\cite{Aubin:2004fs,Bernard:2006wx}.  

For the case of 2+1 flavors, baryons with continuum masses that are unphysical have been identified with partially quenched states.  The fact that the baryons with unphysical masses can all be identified with partially quenched states underscores the conclusion noted in~\cite{Aubin:2003uc,Sharpe:2006re}:  At nonzero lattice spacing, staggered QCD is a necessarily partially quenched theory.  

The analysis leading to Tables~\ref{continuum}, \ref{os1}, and \ref{os2} has also been performed for the case of nondegenerate valence quarks (1+1+1 flavor simulations), but the details are much the same and are therefore omitted here.  In this case the continuum valence symmetry is $SU(4)_x\times SU(4)_y\times SU(4)_z$, and the lattice valence symmetry, $U(1)_x\times U(1)_y\times U(1)_z\times\mr{GTS}$.  The analysis shows that operators with definite isospin (or definite $SU(3)_F$ quantum numbers) do not possess any clear advantage over operators transforming in irreps of $U(1)_x\times U(1)_y\times U(1)_z\times\mr{GTS}$ for interpolating to the non-nucleon states in the ground state multiplet, viz., the $\Sigma$, $\Lambda$, $\Xi$, and $\Lambda_s$.  

\begin{acknowledgments}
C. Bernard has my sincere thanks for suggesting the study of staggered baryons and for much assistance and insightful advice along the way.  I would also like to thank C.~\mbox{DeTar}, D.~Toussaint, Steven~Gottlieb, R.~Sugar, and Dru~Renner for helpful and encouraging discussions.  Funding was supplied in part by the U.S. Department of Energy under grant DE-FG02-91ER40628.
\end{acknowledgments}

\appendix*
\section{\label{excite}Excited baryons and \sugts\ operators}
Here we consider how to extend the analyses of Sec.'s~\ref{spec} and~\ref{operspec} to excited baryons.  As for the ground state baryons, we seek to learn as much as possible about the staggered spectrum by extending and then breaking the valence symmetry group to account respectively for the presence of taste and the violation of continuum taste symmetry.  

In the standard nonrelativistic quark model, the light-quark baryons transform in $SU(6)\times O(3)$ supermultiplets that fall into harmonic oscillator energy bands.  The ground state representation has zero orbital angular momentum, positive parity, and corresponds to the ground state of the oscillator; the ground state baryons transform in $(\mb{56_S},\ 0_0^+)$, where the energy quanta of the oscillator and the parity of the representation are denoted respectively by a subscript and superscript on the orbital angular momentum quantum number.  The first excited energy band of the oscillator contains a $(\mb{70_M},\ 1_1^-)$, and the second energy band, five supermultiplets:  $(\mb{56_S},\ 0_2^+)$, $(\mb{70_M},\ 0_2^+)$, $(\mb{56_S},\ 2_2^+)$, $(\mb{70_M},\ 2_2^+)$, and $(\mb{20_A},\ 1_2^+)$.  Identification of the members of these multiplets with the observed baryons indicates that the states of the $(\mb{56_S},\ 0_2^+)$ are almost always lighter than the corresponding octet and decuplet members of the $(\mb{70_M},\ 1_1^-)$~\cite{PDBook}.  

Extending the quark model to account for the valence symmetry of staggered QCD is straightforward.  One must simply replace the $SU(6)$ multiplets with the corresponding $SU(24)$ multiplets and execute the appropriate decompositions to identify continuum irreps that are degenerate with physical baryons and the lattice irreps of operators that couple to them.  In $SU(6)$ we have \[\mb{6\otimes6\otimes6\rightarrow56_S\oplus70_M\oplus70_M\oplus20_A}.\]  The corresponding decomposition in $SU(24)$ is \[\mb{24\otimes24\otimes24\rightarrow2600_S\oplus4600_M\oplus4600_M\oplus2024_A}.\]  Decomposing the $SU(6)$ irreps under the spin-flavor $SU(2)_S\times SU(3)_F$ gives
\begin{subequations}\label{6to2x3}
\begin{eqnarray}
\mb{56_S}&\rightarrow&\mb{(\thf,\ 10_S)\oplus(\ohf,\ 8_M)}\\
\mb{70_M}&\rightarrow&\mb{(\ohf,\ 10_S)\oplus(\thf,\ 8_M)\oplus(\ohf,\ 8_M)\oplus(\ohf,\ 1_A)}\\
\mb{20_A}&\rightarrow&\mb{(\ohf,\ 8_M)\oplus(\thf,\ 1_A)}.
\end{eqnarray}
\end{subequations}
The corresponding decompositions of $SU(24)$ irreps under $SU(2)_S\times SU(12)_f$ are
\begin{subequations}\label{24to2x12}
\begin{eqnarray}
\mb{2600_S}&\rightarrow&\mb{(\thf,\ 364_S)\oplus(\ohf,\ 572_M)}\label{2600to2x12}\\
\mb{4600_M}&\rightarrow&\mb{(\ohf,\ 364_S)\oplus(\thf,\ 572_M)\oplus(\ohf,\ 572_M)\oplus(\ohf,\ 220_A)}\label{4600to2x12}\\
\mb{2024_A}&\rightarrow&\mb{(\ohf,\ 572_M)\oplus(\thf,\ 220_A)}\label{2024to2x12}.
\end{eqnarray}
\end{subequations}
We classify the staggered baryons in supermultiplets of $SU(24)\times O(3)$ that fall into harmonic oscillator energy bands.  The zeroth energy band contains only the ground state baryons, which transform in the $(\mb{2600_S},\ 0_0^+)$.  The first energy band contains only the negative parity $(\mb{4600_M},\ 1_1^-)$ baryons.  The second energy band contains the five supermultiplets $(\mb{2600_S},\ 0_2^+)$, $(\mb{4600_M},\ 0_2^+)$, $(\mb{2600_S},\ 2_2^+)$, $(\mb{4600_M},\ 2_2^+)$, and $(\mb{2024_A},\ 1_2^+)$.  

The next step is to identify baryons degenerate with physical baryons in the continuum limit; for definiteness, consider the second case of Table~\ref{cases}.  We decompose the $SU(12)_f$ irreps under flavor-taste $SU(3)_F\times SU(4)_T$ and search for symmetric irreps of taste $SU(4)_T$:
\begin{subequations}\label{12to3x4}
\begin{eqnarray}
\mb{364_S}&\rightarrow&\mb{(10_S,\ 20_S)\oplus(8_M,\ 20_M)\oplus(1_A,\ \bar 4_A)}\label{364to3x4}\\
\mb{572_M}&\rightarrow&\mb{(10_S,\ 20_M)\oplus(8_M,\ 20_S)\oplus(8_M,\ 20_M)\oplus(8_M,\ \bar 4_A)\oplus(1_A,\ 20_M)}\label{572to3x4}\\
\mb{220_A}&\rightarrow&\mb{(10_S,\ \bar 4_A)\oplus(8_M,\ 20_M)\oplus(1_A,\ 20_S)}.\label{220to3x4}
\end{eqnarray}
\end{subequations}
In each of these decompositions, the symmetric taste irrep appears with the flavor $SU(3)_F$ irrep that has the same symmetry as the $SU(12)_f$ irrep.  The symmetry of each $SU(12)_f$ irrep appearing in the decompositions (\ref{24to2x12}) is the same as the symmetry of the corresponding $SU(3)_F$ irreps appearing in the decompositions (\ref{6to2x3}).  Finally, the symmetry of each $SU(24)$ irrep in the $SU(24)\times O(3)$ supermultiplets is the same as the symmetry of the $SU(6)$ irrep in the corresponding $SU(6)\times O(3)$ supermultiplets.  Therefore, the fact that the $SU(3)_F$ octet, decuplet, and singlet appear once and only once with the symmetric taste irrep in the decompositions (\ref{12to3x4}) implies that for every resonance identified within the context of the $SU(6)\times O(3)$ quark model, there exists a taste-symmetric $SU(3)_F\times SU(4)_T$ irrep that contains 20 baryons that become degenerate, in the continuum limit, with the given resonance.  

For example, for the ground state supermultiplet, the decompositions (\ref{2600to2x12}) and (\ref{12to3x4}) imply the existence of a spin-\ohf\ octet and a spin-\thf\ decuplet that are respectively degenerate with the lightest octet and decuplet of nature.  The lightest excited states must reside in the $(\mb{2600_S},\ 0_2^+)$, and (\ref{2600to2x12}) then implies the existence of a spin-\ohf\ octet including a state degenerate with the $N(1440)$ and a spin-\thf\ decuplet whose lightest members are degenerate with the $\Delta(1600)$.  Supermultiplets with nontrivial orbital angular momentum that transform in the \mb{2600_S} of $SU(24)$ are handled in much the same way.  Referring to~\cite{PDBook}, we see that the $(\mb{2600_S},\ 2_2^+)$ contains a spin-\thf\ octet containing the $N(1720)$, a spin-\fhf\ octet containing the $N(1680)$, a spin-\fhf\ decuplet containing the $\Delta(1905)$, and so on.  For the $(\mb{4600_M},\ 1_1^-)$, the decompositions (\ref{4600to2x12}) and (\ref{12to3x4}) reveal octets containing the $N(1535)$, $N(1520)$, $N(1650)$, $N(1700)$, and $N(1675)$ and decuplets containing the $\Delta(1620)$ and $\Delta(1700)$ (cf.~\cite{PDBook}).

To identify all the staggered baryons degenerate with the states of these quark-model assignments, we identify the continuum valence symmetry irreps in which the taste-symmetric baryons transform.  Decomposing the $SU(12)_f$ irreps under $SU(8)_{x,y}\times SU(4)_z$ gives
\begin{subequations}\label{12to8x4}
\begin{eqnarray}
\mb{364_S}&\rightarrow&\mb{(120_S,\ 1)\oplus(36_S,\ 4)\oplus(8,\ 10_S)\oplus(1,\ 20_S)},\label{364to8x4}\\
\mb{572_M}&\rightarrow&\mb{(168_M,\ 1)\oplus(28_A,\ 4)\oplus(36_S,\ 4)}\nonumber \\
&&\oplus\;\mb{(8,\ 6_A)\oplus(8,\ 10_S)\oplus(1,\ 20_M)}\label{572to8x4}\\
\mb{220_A}&\rightarrow&\mb{(56_A,\ 1)\oplus(28_A,\ 4)\oplus(8,\ 6_A)\oplus(1,\ \bar 4_A)}.\label{220to8x4}
\end{eqnarray}
\end{subequations}
The decompositions under $SU(2)_I\times SU(4)_T$ of the $SU(8)_{x,y}\times SU(4)_z$ irreps appearing in (\ref{364to8x4}) and (\ref{572to8x4}) are given in (\ref{su842isot}); the decompositions under $SU(2)_I\times SU(4)_T$ of the $SU(3)_F\times SU(4)_T$ irreps appearing in (\ref{364to3x4}) and (\ref{572to3x4}) are given in (\ref{ft2isot}).  For the additional $SU(8)_{x,y}\times SU(4)_z$ irreps appearing in (\ref{220to8x4}), decomposing under $SU(2)_I\times SU(4)_T$ gives
\begin{eqnarray*}
\mb{(56_A,\ 1)}_0&\rightarrow&\mb{(\thf,\ \bar 4_A)}_0\oplus\mb{(\ohf,\ 20_M)}_0\\
\mb{(1,\ \bar 4_A)}_{-3}&\rightarrow&\mb{(0,\ \bar 4_A)}_{-3},
\end{eqnarray*}
while decomposing the new $SU(3)_F\times SU(4)_T$ irreps appearing in (\ref{220to3x4}) gives
\begin{eqnarray*}
\mb{(10_S,\ \bar 4_A)}&\rightarrow&\mb{(\thf,\ \bar 4_A)}_0\oplus\mb{(1,\ \bar 4_A)}_{-1}\oplus\mb{(\ohf,\ \bar 4_A)}_{-2}\oplus\mb{(0,\ \bar 4_A)}_{-3}\\
\mb{(1_A,\ 20_S)}&\rightarrow&\mb{(0,\ 20_S)}_{-1}.
\end{eqnarray*}
Noting that the decompositions of the \mb{364_S} and \mb{572_M} are the same for excited supermultiplets as for the ground state supermultiplet, we see that the first three columns of Table~\ref{continuum} and the observations summarized in~(\ref{switch}) can be immediately taken over for the excited states.  

To identify continuum valence irreps of baryons degenerate with the physical $SU(3)_F$ singlets, we search for $\mb{(0,\ 20_S)}_{-1}$ irreps in the decompositions of the irreps in (\ref{220to8x4}) under $SU(2)_I\times SU(4)_T$.  Because only the \mb{(28_A,\ 4)} contains such an irrep, we conclude that the members of the \mb{(28_A,\ 4)} become degenerate with the singlets of the corresponding physical supermultiplets.  Under interchange of the valence up-down and strange quark masses, the masses of the \mb{(28_A,\ 4)} and \mb{(8,\ 6_A)} are interchanged, as are the masses of the \mb{(56_A,\ 1)} and \mb{(1,\ \bar 4_A)}.  As for continuum irreps corresponding to physical octets, the \mb{(28_A,\ 4)} corresponds to the $\Lambda$, and the \mb{(8,\ 6_A)}, to the $\Lambda_s$.  The \mb{(56_A,\ 1)} and \mb{(1,\ \bar 4_A)} are not constrained to have physical masses in the continuum limit; in what follows, they are respectively denoted the $\Lambda_u$ and $\Lambda_{ss}$ irreps.  Expected continuum degeneracies for some of the baryons in the $(\mb{2600_S},\ 0_2^+)$, $(\mb{4600_M},\ 1_1^-)$, $(\mb{4600_M},\ 0_2^+)$, and $(\mb{2600_S},\ 2_2^+)$ are given in Tables~\ref{excont1},~\ref{excont2a},~\ref{excont2b},~\ref{excont2c},~\ref{excont3}, and~\ref{excont4}.  The degeneracies correspond to quark-model assignments and therefore inherit the limitations of the quark model; mixings between baryons with the same spin and parity are often quite large, and some assignments are highly tentative~\cite{PDBook}.  
\begin{table}
\begin{ruledtabular}
\begin{tabular}{llll}
Case and symmetry for $a=0$ & Irreps for $a=0$ & Expected mass & No. of lattice irreps \\ \hline
$m_x=m_y=m_z=\hat m$ & \mb{(\ohf,\ 572_M)} & $N(1440)$ & 16\\
$\su{2}{S}\times\su{12}{f}$ & \mb{(\thf,\ 364_S)} & $\Delta(1600)$ & 14\\ \hline
$m_x=m_y=\hat m,\;m_z=m_s$ & \mb{(\ohf,\ 168_M,\ 1)} & $N(1440)$ & 12\\
$\su{2}{S}\times\su{8}{x,y}\times\su{4}{z}$ & \mb{(\ohf,\ 28_A,\ 4)} & $\Lambda(1600)$ & 12\\
 & \mb{(\ohf,\ 36_S,\ 4)} & $\Sigma(1660)$ & 12\\
 & \mb{(\ohf,\ 8,\ 10_S)} & $\Xi(?)$ & 7\\
 & \mb{(\ohf,\ 8,\ 6_A)} & $\Lambda_s(?)$ & 5\\
 & \mb{(\ohf,\ 1,\ 20_M)} & $N_s(?)$ & 4\\
 & \mb{(\thf,\ 120_S,\ 1)} & $\Delta(1600)$ & 13\\
 & \mb{(\thf,\ 36_S,\ 4)} & $\Sigma^*(?)$ & 20\\
 & \mb{(\thf,\ 8,\ 10_S)} & $\Xi^*(?)$ & 13\\
 & \mb{(\thf,\ 1,\ 20_S)} & $\Omega^-(?)$ & 7
\end{tabular}
\end{ruledtabular}
\caption{\label{excont1}For baryons in the $(\mb{2600_S},\ 0_2^+)$, expected continuum degeneracies and the number of corresponding lattice irreps.  The results are qualitatively identical to the results of Table~\ref{continuum}.}
\end{table}
\begin{table}
\begin{ruledtabular}
\begin{tabular}{llll}
Case and symmetry for $a=0$ & Irreps for $a=0$ & Expected mass & No. of lattice irreps \\ \hline
$m_x=m_y=m_z=\hat m$ & \mb{(\ohf,\ 572_M)} & $N(1535)$ & 16\\
$\su{2}{J}\times\su{12}{f}$ & \mb{(\thf,\ 572_M)} & $N(1520)$ & 26\\ \hline
$m_x=m_y=\hat m,\;m_z=m_s$ & \mb{(\ohf,\ 168_M,\ 1)} & $N(1535)$ & 12\\
$\su{2}{J}\times\su{8}{x,y}\times\su{4}{z}$ & \mb{(\ohf,\ 28_A,\ 4)} & $\Lambda(1670)$ & 12\\
 & \mb{(\ohf,\ 36_S,\ 4)} & $\Sigma(1620)$ & 12\\
 & \mb{(\ohf,\ 8,\ 10_S)} & $\Xi(?)$ & 7\\
 & \mb{(\ohf,\ 8,\ 6_A)} & $\Lambda_s(?)$ & 5\\
 & \mb{(\ohf,\ 1,\ 20_M)} & $N_s(?)$ & 4\\
 & \mb{(\thf,\ 168_M,\ 1)} & $N(1520)$ & 20\\
 & \mb{(\thf,\ 28_A,\ 4)} & $\Lambda(1690)$ & 20\\
 & \mb{(\thf,\ 36_S,\ 4)} & $\Sigma(1670)$ & 20\\
 & \mb{(\thf,\ 8,\ 10_S)} & $\Xi(1820)$ & 13\\
 & \mb{(\thf,\ 8,\ 6_A)} & $\Lambda_s(?)$ & 7\\
 & \mb{(\thf,\ 1,\ 20_M)} & $N_s(?)$ & 6
\end{tabular}
\end{ruledtabular}
\caption{\label{excont2a}For baryons in the \mb{(\ohf,\ 572_M)} in $(\mb{4600_M},\ 1_1^-)$ (cf.~(\ref{4600to2x12})), expected continuum degeneracies and the number of corresponding lattice irreps.  Note that the spin $J$ of the baryon is no longer equal to the net spin $S$ of the quarks.}
\end{table}
\begin{table}
\begin{ruledtabular}
\begin{tabular}{llll}
Case and symmetry for $a=0$ & Irreps for $a=0$ & Expected mass & No. of lattice irreps \\ \hline
$m_x=m_y=m_z=\hat m$ & \mb{(\ohf,\ 364_S)} & $\Delta(1620)$ & 8\\
$\su{2}{J}\times\su{12}{f}$ & \mb{(\thf,\ 364_S)} & $\Delta(1700)$ & 14\\
& \mb{(\ohf,\ 220_A)} & $\Lambda_u(?)$ & 8\\ 
& \mb{(\thf,\ 220_A)} & $\Lambda_u(?)$ & 14\\ \hline
$m_x=m_y=\hat m,\;m_z=m_s$  & \mb{(\ohf,\ 120_S,\ 1)} & $\Delta(1620)$ & 7\\
$\su{2}{J}\times\su{8}{x,y}\times\su{4}{z}$  & \mb{(\ohf,\ 36_S,\ 4)} & $\Sigma^*(?)$ & 12\\
 & \mb{(\ohf,\ 8,\ 10_S)} & $\Xi^*(?)$ & 7\\
 & \mb{(\ohf,\ 1,\ 20_S)} & $\Omega^-(?)$ & 3\\
 & \mb{(\thf,\ 120_S,\ 1)} & $\Delta(1700)$ & 13\\
 & \mb{(\thf,\ 36_S,\ 4)} & $\Sigma^*(?)$ & 20\\
 & \mb{(\thf,\ 8,\ 10_S)} & $\Xi^*(?)$ & 13\\
 & \mb{(\thf,\ 1,\ 20_S)} & $\Omega^-(?)$ & 7\\
 & \mb{(\ohf,\ 56_A,\ 1)} & $\Lambda_u(?)$ & 5\\
 & \mb{(\ohf,\ 28_A,\ 4)} & $\Lambda(1405)$ & 12\\
 & \mb{(\ohf,\ 8,\ 6_A)} & $\Lambda_s(?)$ & 5\\
 & \mb{(\ohf,\ 1,\ \bar 4_A)} & $\Lambda_{ss}(?)$ & 1\\
 & \mb{(\thf,\ 56_A,\ 1)} & $\Lambda_u(?)$ & 7\\
 & \mb{(\thf,\ 28_A,\ 4)} & $\Lambda(1520)$ & 20\\
 & \mb{(\thf,\ 8,\ 6_A)} & $\Lambda_s(?)$ & 7\\
 & \mb{(\thf,\ 1,\ \bar 4_A)} & $\Lambda_{ss}(?)$ & 1
\end{tabular}
\end{ruledtabular}
\caption{\label{excont2b}For baryons in the \mb{(\ohf,\ 364_S)} and \mb{(\ohf,\ 220_A)} in $(\mb{4600_M},\ 1_1^-)$ (cf.~(\ref{4600to2x12})), expected continuum degeneracies and the number of corresponding lattice irreps.}
\end{table}
\begin{table}
\begin{ruledtabular}
\begin{tabular}{llll}
Case and symmetry for $a=0$ & Irreps for $a=0$ & Expected mass & No. of lattice irreps \\ \hline
$m_x=m_y=m_z=\hat m$ & \mb{(\ohf,\ 572_M)} & $N(1650)$ & 16\\
$\su{2}{J}\times\su{12}{f}$ & \mb{(\thf,\ 572_M)} & $N(1700)$ & 26\\ 
 & \mb{(\fhf,\ 572_M)} & $N(1675)$ & 42\\ \hline
$m_x=m_y=\hat m,\;m_z=m_s$ & \mb{(\ohf,\ 168_M,\ 1)} & $N(1650)$ & 12\\
$\su{2}{J}\times\su{8}{x,y}\times\su{4}{z}$ & \mb{(\ohf,\ 28_A,\ 4)} & $\Lambda(1800)$ & 12\\
 & \mb{(\ohf,\ 36_S,\ 4)} & $\Sigma(1750)$ & 12\\
 & \mb{(\ohf,\ 8,\ 10_S)} & $\Xi(?)$ & 7\\
 & \mb{(\ohf,\ 8,\ 6_A)} & $\Lambda_s(?)$ & 5\\
 & \mb{(\ohf,\ 1,\ 20_M)} & $N_s(?)$ & 4\\
 & \mb{(\thf,\ 168_M,\ 1)} & $N(1700)$ & 20\\
 & \mb{(\thf,\ 28_A,\ 4)} & $\Lambda(?)$ & 20\\
 & \mb{(\thf,\ 36_S,\ 4)} & $\Sigma(?)$ & 20\\
 & \mb{(\thf,\ 8,\ 10_S)} & $\Xi(?)$ & 13\\
 & \mb{(\thf,\ 8,\ 6_A)} & $\Lambda_s(?)$ & 7\\
 & \mb{(\thf,\ 1,\ 20_M)} & $N_s(?)$ & 6\\
 & \mb{(\fhf,\ 168_M,\ 1)} & $N(1675)$ & 32\\
 & \mb{(\fhf,\ 28_A,\ 4)} & $\Lambda(1830)$ & 32\\
 & \mb{(\fhf,\ 36_S,\ 4)} & $\Sigma(1775)$ & 32\\
 & \mb{(\fhf,\ 8,\ 10_S)} & $\Xi(?)$ & 20\\
 & \mb{(\fhf,\ 8,\ 6_A)} & $\Lambda_s(?)$ & 12\\
 & \mb{(\fhf,\ 1,\ 20_M)} & $N_s(?)$ & 10
\end{tabular}
\end{ruledtabular}
\caption{\label{excont2c}For baryons in the \mb{(\thf,\ 572_M)} in $(\mb{4600_M},\ 1_1^-)$ (cf.~(\ref{4600to2x12})), expected continuum degeneracies and the number of corresponding lattice irreps.}
\end{table}
\begin{table}
\begin{ruledtabular}
\begin{tabular}{llll}
Case and symmetry for $a=0$ & Irreps for $a=0$ & Expected mass & No. of lattice irreps \\ \hline
$m_x=m_y=m_z=\hat m$ & \mb{(\ohf,\ 572_M)} & $N(1710)$ & 16\\
$\su{2}{S}\times\su{12}{f}$ &  & & \\ \hline
$m_x=m_y=\hat m,\;m_z=m_s$ & \mb{(\ohf,\ 168_M,\ 1)} & $N(1710)$ & 12\\
$\su{2}{S}\times\su{8}{x,y}\times\su{4}{z}$ & \mb{(\ohf,\ 28_A,\ 4)} & $\Lambda(1810)$ & 12\\
 & \mb{(\ohf,\ 36_S,\ 4)} & $\Sigma(1880)$ & 12\\
 & \mb{(\ohf,\ 8,\ 10_S)} & $\Xi(?)$ & 7\\
 & \mb{(\ohf,\ 8,\ 6_A)} & $\Lambda_s(?)$ & 5\\
 & \mb{(\ohf,\ 1,\ 20_M)} & $N_s(?)$ & 4
\end{tabular}
\end{ruledtabular}
\caption{\label{excont3}For baryons in the \mb{(\ohf,\ 572_M)} in $(\mb{4600_M},\ 0_2^+)$ (cf.~(\ref{4600to2x12})), expected continuum degeneracies and the number of corresponding lattice irreps.  The results are qualitatively identical to the results of Table~\ref{continuum}.}
\end{table}
\begin{table}
\begin{ruledtabular}
\begin{tabular}{llll}
Case and symmetry for $a=0$ & Irreps for $a=0$ & Expected mass & No. of lattice irreps \\ \hline
$m_x=m_y=m_z=\hat m$ & \mb{(\thf,\ 572_M)} & $N(1720)$ & 26\\
$\su{2}{J}\times\su{12}{f}$ & \mb{(\fhf,\ 572_M)} & $N(1680)$ & 42\\
 & \mb{(\fhf,\ 364_S)} & $\Delta(1905)$ & 22\\
 & \mb{(\shf,\ 364_S)} & $\Delta(1950)$ & 30\\ \hline
$m_x=m_y=\hat m,\;m_z=m_s$ & \mb{(\thf,\ 168_M,\ 1)} & $N(1720)$ & 20\\
$\su{2}{J}\times\su{8}{x,y}\times\su{4}{z}$ & \mb{(\thf,\ 28_A,\ 4)} & $\Lambda(1890)$ & 20\\
 & \mb{(\thf,\ 36_S,\ 4)} & $\Sigma(?)$ & 20\\
 & \mb{(\thf,\ 8,\ 10_S)} & $\Xi(?)$ & 13\\
 & \mb{(\thf,\ 8,\ 6_A)} & $\Lambda_s(?)$ & 7\\
 & \mb{(\thf,\ 1,\ 20_M)} & $N_s(?)$ & 6\\
 & \mb{(\fhf,\ 168_M,\ 1)} & $N(1680)$ & 32\\
 & \mb{(\fhf,\ 28_A,\ 4)} & $\Lambda(1820)$ & 32\\
 & \mb{(\fhf,\ 36_S,\ 4)} & $\Sigma(1915)$ & 32\\
 & \mb{(\fhf,\ 8,\ 10_S)} & $\Xi(2030)$ & 20\\
 & \mb{(\fhf,\ 8,\ 6_A)} & $\Lambda_s(?)$ & 12\\
 & \mb{(\fhf,\ 1,\ 20_M)} & $N_s(?)$ & 10\\
 & \mb{(\fhf,\ 120_S,\ 1)} & $\Delta(1905)$ & 20\\
 & \mb{(\fhf,\ 36_S,\ 4)} & $\Sigma^*(?)$ & 32\\
 & \mb{(\fhf,\ 8,\ 10_S)} & $\Xi^*(?)$ & 20\\
 & \mb{(\fhf,\ 1,\ 20_S)} & $\Omega^-(?)$ & 10\\
 & \mb{(\shf,\ 120_S,\ 1)} & $\Delta(1950)$ & 27\\
 & \mb{(\shf,\ 36_S,\ 4)} & $\Sigma^*(2030)$ & 44\\
 & \mb{(\shf,\ 8,\ 10_S)} & $\Xi^*(?)$ & 27\\
 & \mb{(\shf,\ 1,\ 20_S)} & $\Omega^-(?)$ & 13
\end{tabular}
\end{ruledtabular}
\caption{\label{excont4}For baryons in the \mb{(\ohf,\ 572_M)} and the $J=\fhf,\ \shf$ members of the \mb{(\thf,\ 364_S)} in $(\mb{2600_S},\ 2_2^+)$, expected continuum degeneracies and the number of corresponding lattice irreps.}
\end{table}

At nonzero lattice spacing, we consider the decomposition of the continuum $\mc{P}\times SU(2)_J\times SU(4)_T$ irreps under $\mc{P}\times\gts$.  For $J=\ohf,\ \thf$ the relevant decompositions (suppressing parity quantum numbers) are given in (\ref{taste2gts}).  For $J=\fhf,\ \shf$ we also require
\begin{subequations}
\label{taste2gtsb}
\begin{eqnarray}
SU(2)_J\times SU(4)_T&\supset& \mr{GTS} \nonumber \\
\mb{(\fhf,\ 20_S)}&\rightarrow& 2\mb{(8)}\oplus3\mb{(8^{\prime})}\oplus5\mb{(16)}\label{t1b}\\
\mb{(\fhf,\ 20_M)}&\rightarrow& \mb{8}\oplus4\mb{(8^{\prime})}\oplus5\mb{(16)}\label{t2b}\\
\mb{(\fhf,\ \bar 4_A)}&\rightarrow& \mb{8^{\prime}\oplus16}\label{t3b}\\
\mb{(\shf,\ 20_S)}&\rightarrow& 3\mb{(8)}\oplus3\mb{(8^{\prime})}\oplus7\mb{(16)}\label{t4b}\\
\mb{(\shf,\ 20_M)}&\rightarrow& 4\mb{(8)}\oplus4\mb{(8^{\prime})}\oplus6\mb{(16)}\label{t5b}\\
\mb{(\shf,\ \bar 4_A)}&\rightarrow& \mb{8}\oplus\mb{8^{\prime}}\oplus\mb{16}.\label{t6b}
\end{eqnarray}
\end{subequations}
Together with the decompositions (\ref{ks2gts}), (\ref{su2su842isosgts1}), and (\ref{su2su842isosgts2}), we find
\begin{subequations}
\label{eks2gts}
\begin{eqnarray}
SU(2)_J\times SU(12)_f&\supset& \sugts \nonumber\\
\mb{(\ohf,\ 364_S)}&\rightarrow& \mb{(10_S,\ 8)}\oplus2\mb{(10_S,\ 16)}\oplus3\mb{(8_M,\ 8)}\oplus\mb{(8_M,\ 16)}\oplus\mb{(1_A,\ 8)}\label{e364}\\
\mb{(\fhf,\ 364_S)}&\rightarrow& 2\mb{(10_S,\ 8)}\oplus3\mb{(10_S,\ 8^{\prime})}\oplus5\mb{(10_S,\ 16)}\oplus\mb{(8_M,\ 8)}\nonumber\\
&&\oplus\;4\mb{(8_M,\ 8^{\prime})}\oplus5\mb{(8_M,\ 16)}\oplus\mb{(1_A,\ 8^{\prime})}\oplus\mb{(1_A,\ 16)}\label{e364a}\\
\mb{(\shf,\ 364_S)}&\rightarrow& 3\mb{(10_S,\ 8)}\oplus3\mb{(10_S,\ 8^{\prime})}\oplus7\mb{(10_S,\ 16)}\oplus4\mb{(8_M,\ 8)}\oplus4\mb{(8_M,\ 8^{\prime})}\nonumber\\
&&\oplus\;6\mb{(8_M,\ 16)}\oplus\mb{(1_A,\ 8)}\oplus\mb{(1_A,\ 8^{\prime})}\oplus\mb{(1_A,\ 16)}\label{e364b}\\
\mb{(\thf,\ 572_M)}&\rightarrow& \mb{(10_S,\ 8)}\oplus\mb{(10_S,\ 8^{\prime})}\oplus4\mb{(10_S,\ 16)}\oplus3\mb{(8_M,\ 8)}\oplus3\mb{(8_M,\ 8^{\prime})}\nonumber\\
&&\oplus\;8\mb{(8_M,\ 16)}\oplus\mb{(1_A,\ 8)\oplus(1_A,\ 8^{\prime})}\oplus4\mb{(1_A,\ 16)}\label{e572a}\\
\mb{(\fhf,\ 572_M)}&\rightarrow& \mb{(10_S,\ 8)}\oplus4\mb{(10_S,\ 8^{\prime})}\oplus5\mb{(10_S,\ 16)}\oplus3\mb{(8_M,\ 8)}\oplus8\mb{(8_M,\ 8^{\prime})}\nonumber\\
&&\oplus\;11\mb{(8_M,\ 16)}\oplus\mb{(1_A,\ 8)}\oplus4\mb{(1_A,\ 8^{\prime})}\oplus5\mb{(1_A,\ 16)}\label{e572b}\\
\mb{(\ohf,\ 220_A)}&\rightarrow& \mb{(10_S,\ 8)}\oplus3\mb{(8_M,\ 8)}\oplus\mb{(8_M,\ 16)}\oplus\mb{(1_A,\ 8)}\oplus2\mb{(1_A,\ 16)}\label{e220a}\\
\mb{(\thf,\ 220_A)}&\rightarrow& \mb{(10_S,\ 16)}\oplus\mb{(8_M,\ 8)}\oplus\mb{(8_M,\ 8^{\prime})}\oplus4\mb{(8_M,\ 16)}\nonumber\\
&&\oplus\;2\mb{(1_A,\ 8)}\oplus2\mb{(1_A,\ 8^{\prime})}\oplus3\mb{(1_A,\ 16)},\label{e220b}
\end{eqnarray}
\end{subequations}
\begin{subequations}
\label{esu2su842isosgts1}
\begin{eqnarray}
SU(2)_J\;\,\times\;\,SU(8)_{x,y}&\times&SU(4)_z\;\,\supset\;\,SU(2)_I\;\,\times\;\,\mr{GTS} \nonumber\\
\mb{(\ohf,\ 120_S,\ 1)}&\rightarrow&\mb{(\thf,\ 8)}_0\oplus2\mb{(\thf,\ 16)}_0\oplus3\mb{(\ohf,\ 8)}_0\oplus\mb{(\ohf,\ 16)}_0\label{eg1}\\
\mb{(\ohf,\ 1,\ 20_S)}&\rightarrow&\mb{(0,\ 8)}_{-3}\oplus2\mb{(0,\ 16)}_{-3}\label{eg4}\\
\mb{(\fhf,\ 120_S,\ 1)}&\rightarrow&2\mb{(\thf,\ 8)}_0\oplus3\mb{(\thf,\ 8^{\prime})}_0\oplus5\mb{(\thf,\ 16)}_0\nonumber\\
&&\oplus\;\mb{(\ohf,\ 8)}_0\oplus4\mb{(\ohf,\ 8^{\prime})}_0\oplus5\mb{(\ohf,\ 16)}_0\label{eg1b}\\
\mb{(\fhf,\ 36_S,\ 4)}&\rightarrow&3\mb{(1,\ 8)}_{-1}\oplus7\mb{(1,\ 8^{\prime})}_{-1}\oplus10\mb{(1,\ 16)}_{-1}\nonumber\\
&&\oplus\;\mb{(0,\ 8)}_{-1}\oplus5\mb{(0,\ 8^{\prime})}_{-1}\oplus6\mb{(0,\ 16)}_{-1}\label{eg2b}\\
\mb{(\fhf,\ 8,\ 10_S)}&\rightarrow&3\mb{(\ohf,\ 8)}_{-2}\oplus7\mb{(\ohf,\ 8^{\prime})}_{-2}\oplus10\mb{(\ohf,\ 16)}_{-2}\label{eg3b}\\
\mb{(\fhf,\ 1,\ 20_S)}&\rightarrow&2\mb{(0,\ 8)}_{-3}\oplus3\mb{(0,\ 8^{\prime})}_{-3}\oplus5\mb{(0,\ 16)}_{-3}\label{eg4b}\\
\mb{(\shf,\ 120_S,\ 1)}&\rightarrow&3\mb{(\thf,\ 8)}_0\oplus3\mb{(\thf,\ 8^{\prime})}_0\oplus7\mb{(\thf,\ 16)}_0\nonumber\\
&&\oplus\;4\mb{(\ohf,\ 8)}_0\oplus4\mb{(\ohf,\ 8^{\prime})}_0\oplus6\mb{(\ohf,\ 16)}_0\label{eg1c}\\
\mb{(\shf,\ 36_S,\ 4)}&\rightarrow&7\mb{(1,\ 8)}_{-1}\oplus7\mb{(1,\ 8^{\prime})}_{-1}\oplus13\mb{(1,\ 16)}_{-1}\nonumber\\
&&\oplus\;5\mb{(0,\ 8)}_{-1}\oplus5\mb{(0,\ 8^{\prime})}_{-1}\oplus7\mb{(0,\ 16)}_{-1}\label{eg2c}\\
\mb{(\shf,\ 8,\ 10_S)}&\rightarrow&7\mb{(\ohf,\ 8)}_{-2}\oplus7\mb{(\ohf,\ 8^{\prime})}_{-2}\oplus13\mb{(\ohf,\ 16)}_{-2}\label{eg3c}\\
\mb{(\shf,\ 1,\ 20_S)}&\rightarrow&3\mb{(0,\ 8)}_{-3}\oplus3\mb{(0,\ 8^{\prime})}_{-3}\oplus7\mb{(0,\ 16)}_{-3},\label{eg4c}
\end{eqnarray}
\end{subequations}
\begin{subequations}
\label{esu2su842isosgts2}
\begin{eqnarray}
SU(2)_J\;\,\times\;\,SU(8)_{x,y}&\times&SU(4)_z\;\,\supset\;\,SU(2)_I\;\,\times\;\,\mr{GTS} \nonumber\\
\mb{(\thf,\ 168_M,\ 1)}&\rightarrow& \mb{(\thf,\ 8)}_0\oplus\mb{(\thf,\ 8^{\prime})}_0\oplus4\mb{(\thf,\ 16)}_0\nonumber\\
&&\oplus\;3\mb{(\ohf,\ 8)}_0\oplus3\mb{(\ohf,\ 8^{\prime})}_0\oplus8\mb{(\ohf,\ 16)}_0\label{eh1}\\
\mb{(\thf,\ 28_A,\ 4)}&\rightarrow& \mb{(1,\ 8)}_{-1}\oplus\mb{(1,\ 8^{\prime})}_{-1}\oplus5\mb{(1,\ 16)}_{-1}\nonumber\\
&&\oplus\;3\mb{(0,\ 8)}_{-1}\oplus3\mb{(0,\ 8^{\prime})}_{-1}\oplus7\mb{(0,\ 16)}_{-1}\label{eh2}\\
\mb{(\thf,\ 8,\ 6_A)}&\rightarrow&\mb{(\ohf,\ 8)}_{-2}\oplus\mb{(\ohf,\ 8^{\prime})}_{-2}\oplus5\mb{(\ohf,\ 16)}_{-2}\label{eh5}\\
\mb{(\thf,\ 1,\ 20_M)}&\rightarrow&\mb{(0,\ 8)}_{-3}\oplus\mb{(0,\ 8^{\prime})}_{-3}\oplus4\mb{(0,\ 16)}_{-3}\label{eh6}\\
\mb{(\fhf,\ 168_M,\ 1)}&\rightarrow&\mb{(\thf,\ 8)}_0\oplus4\mb{(\thf,\ 8^{\prime})}_0\oplus5\mb{(\thf,\ 16)}_0\nonumber\\
&&\oplus\;3\mb{(\ohf,\ 8)}_0\oplus8\mb{(\ohf,\ 8^{\prime})}_0\oplus11\mb{(\ohf,\ 16)}_0\label{eh1b}\\
\mb{(\fhf,\ 28_A,\ 4)}&\rightarrow&\mb{(1,\ 8)}_{-1}\oplus5\mb{(1,\ 8^{\prime})}_{-1}\oplus6\mb{(1,\ 16)}_{-1}\nonumber\\
&&\oplus\;3\mb{(0,\ 8)}_{-1}\oplus7\mb{(0,\ 8^{\prime})}_{-1}\oplus10\mb{(0,\ 16)}_{-1}\label{eh2b}\\
\mb{(\fhf,\ 36_S,\ 4)}&\rightarrow&3\mb{(1,\ 8)}_{-1}\oplus7\mb{(1,\ 8^{\prime})}_{-1}\oplus10\mb{(1,\ 16)}_{-1}\nonumber\\
&&\oplus\;\mb{(0,\ 8)}_{-1}\oplus5\mb{(0,\ 8^{\prime})}_{-1}\oplus6\mb{(0,\ 16)}_{-1}\label{eh3b}\\
\mb{(\fhf,\ 8,\ 10_S)}&\rightarrow&3\mb{(\ohf,\ 8)}_{-2}\oplus7\mb{(\ohf,\ 8^{\prime})}_{-2}\oplus10\mb{(\ohf,\ 16)}_{-2}\label{eh4b}\\
\mb{(\fhf,\ 8,\ 6_A)}&\rightarrow&\mb{(\ohf,\ 8)}_{-2}\oplus5\mb{(\ohf,\ 8^{\prime})}_{-2}\oplus6\mb{(\ohf,\ 16)}_{-2}\label{eh5b}\\
\mb{(\fhf,\ 1,\ 20_M)}&\rightarrow&\mb{(0,\ 8)}_{-3}\oplus4\mb{(0,\ 8^{\prime})}_{-3}\oplus5\mb{(0,\ 16)}_{-3},\label{eh6b}
\end{eqnarray}
\end{subequations}
\begin{subequations}
\label{esu2su842isosgts3}
\begin{eqnarray}
SU(2)_J\;\,\times\;\,SU(8)_{x,y}&\times&SU(4)_z\;\,\supset\;\,SU(2)_I\;\,\times\;\,\mr{GTS} \nonumber\\
\mb{(\ohf,\ 56_A,\ 1)}&\rightarrow& \mb{(\thf,\ 8)}_0\oplus3\mb{(\ohf,\ 8)}_0\oplus\mb{(\ohf,\ 16)}_0\label{eh1c}\\
\mb{(\ohf,\ 1,\ \bar 4_A)}&\rightarrow&\mb{(0,\ 8)}_{-3}\label{eh4c}\\
\mb{(\thf,\ 56_A,\ 1)}&\rightarrow& \mb{(\thf,\ 16)}_0\oplus\mb{(\ohf,\ 8)}_0\oplus\mb{(\ohf,\ 8^{\prime})}_0\oplus4\mb{(\ohf,\ 16)}_0\label{eh1d}\\
\mb{(\thf,\ 28_A,\ 4)}&\rightarrow& \mb{(1,\ 8)}_{-1}\oplus\mb{(1,\ 8^{\prime})}_{-1}\oplus5\mb{(1,\ 16)}_{-1}\nonumber\\
&&\oplus\;3\mb{(0,\ 8)}_{-1}\oplus3\mb{(0,\ 8^{\prime})}_{-1}\oplus7\mb{(0,\ 16)}_{-1}\label{eh2d}\\
\mb{(\thf,\ 8,\ 6_A)}&\rightarrow&\mb{(\ohf,\ 8)}_{-2}\oplus\mb{(\ohf,\ 8^{\prime})}_{-2}\oplus5\mb{(\ohf,\ 16)}_{-2}\label{eh3d}\\
\mb{(\thf,\ 1,\ \bar 4_A)}&\rightarrow&\mb{(0,\ 16)}_{-3}.\label{eh4d}
\end{eqnarray}
\end{subequations}
Counting the number of lattice irreps occuring in each decomposition gives the fourth column in Tables~\ref{excont1},~\ref{excont2a},~\ref{excont2b},~\ref{excont2c},~\ref{excont3}, and~\ref{excont4}.  As for the analogous results for the ground state multiplets (cf.~Table~\ref{continuum}), the results summarized in these tables remain valid if one everywhere makes the replacements $\hat m\leftrightarrow m_s$,~(\ref{switch}), and $\Lambda_u\leftrightarrow\Lambda_{ss}$.  From these tables and the decompositions~(\ref{ks2gts}), (\ref{eks2gts}), (\ref{su2su842isosgts1}), (\ref{su2su842isosgts2}), (\ref{esu2su842isosgts1}), (\ref{esu2su842isosgts2}), and (\ref{esu2su842isosgts3}), we find quark-model assignments of the continuum irreps overlapped by operators transforming irreducibly under \sugts.  These results are given in Tables~\ref{enucos}, \ref{edelos}, \ref{elamos}, \ref{estralamos}, \ref{esigos}, \ref{esigstarandlamos}, \ref{exios}, \ref{existaros}, \ref{estranucos}, and \ref{eomegos}.  Columns for energy levels corresponding to the same type of continuum spin-flavor irrep are identical.  In particular, the column for states that become degenerate with the $N(939)$ would be identical to that for the $N(1440)$, and that for the $\Delta(1232)$ would be identical to that for the $\Delta(1600)$ (see Tables~\ref{enucos} and~\ref{edelos}).
\begin{table}
\begin{ruledtabular}
\begin{tabular}{lccccccccc}
\multicolumn{1}{l}{Operator}&\multicolumn{9}{c}{Excited nucleons created/mixed}\\ 
                         & $N(1440)$ & $N(1520)$ & $N(1535)$ & $N(1650)$ & $N(1675)$ & $N(1680)$ & $N(1700)$ & $N(1710)$ & $N(1720)$ \\
                         & $J^P=\ohf^+$ & $\thf^-$ & $\ohf^-$ & $\ohf^-$ & $\fhf^-$  & $\fhf^+$  & $\thf^-$  & $\ohf^+$  & $\thf^+$  \\ \hline
\mb{(10_S,\ 8)}          & 3         & 1         & 3         & 3         & 1         & 1         & 1         & 3         & 1         \\
\mb{(10_S,\ 8^{\prime})} & -         & 1         & -         & -         & 4         & 4         & 1         & -         & 1         \\
\mb{(10_S,\ 16)}         & 1         & 4         & 1         & 1         & 5         & 5         & 4         & 1         & 4         \\
\mb{(8_M,\ 8)}           & 5         & 3         & 5         & 5         & 3         & 3         & 3         & 5         & 3         \\
\mb{(8_M,\ 8^{\prime})}  & -         & 3         & -         & -         & 8         & 8         & 3         & -         & 3         \\
\mb{(8_M,\ 16)}          & 3         & 8         & 3         & 3         & 11        & 11        & 8         & 3         & 8         \\
\mb{(1_A,\ 8)}           & 3         & 1         & 3         & 3         & 1         & 1         & 1         & 3         & 1         \\
\mb{(1_A,\ 8^{\prime})}  & -         & 1         & -         & -         & 4         & 4         & 1         & -         & 1         \\
\mb{(1_A,\ 16)}          & 1         & 4         & 1         & 1         & 5         & 5         & 4         & 1         & 4
\end{tabular}
\end{ruledtabular}
\caption{\label{enucos}For $m_x=m_y=m_z=\hat m$, the number of nondegenerate staggered states in the dominant continuum irreps of the listed excited nucleons.}
\end{table}
\begin{table}
\begin{ruledtabular}
\begin{tabular}{lccccccc}
\multicolumn{1}{l}{Operator}&\multicolumn{7}{c}{Excited $\Delta$'s and $\Lambda_u$'s created/mixed}\\
                         & $\Delta(1600)$ & $\Delta(1620)$ & $\Delta(1700)$ & $\Delta(1905)$ & $\Delta(1950)$ & $\Lambda_u(?)$ & $\Lambda_u(?)$ \\
                         & $J^P=\thf^+$   & $\ohf^-$       & $\thf^-$       & $\fhf^+$       & $\shf^+$       & $\ohf^-$       & $\thf^-$ \\ \hline
\mb{(10_S,\ 8)}          & 2              & 1              & 2              & 2              & 3              & 1              & - \\
\mb{(10_S,\ 8^{\prime})} & 2              & -              & 2              & 3              & 3              & -              & - \\
\mb{(10_S,\ 16)}         & 3              & 2              & 3              & 5              & 7              & -              & 1 \\
\mb{(8_M,\ 8)}           & 1              & 3              & 1              & 1              & 4              & 3              & 1 \\
\mb{(8_M,\ 8^{\prime})}  & 1              & -              & 1              & 4              & 4              & -              & 1 \\
\mb{(8_M,\ 16)}          & 4              & 1              & 4              & 5              & 6              & 1              & 4 \\
\mb{(1_A,\ 8)}           & -              & 1              & -              & -              & 1              & 1              & 2 \\
\mb{(1_A,\ 8^{\prime})}  & -              & -              & -              & 1              & 1              & -              & 2 \\
\mb{(1_A,\ 16)}          & 1              & -              & 1              & 1              & 1              & 2              & 3
\end{tabular}
\end{ruledtabular}
\caption{\label{edelos}For $m_x=m_y=m_z=\hat m$, the number of nondegenerate staggered states in the dominant continuum irreps of the listed excited $\Delta$'s and $\Lambda_u$'s.}
\end{table}
\begin{table}
\begin{ruledtabular}
\begin{tabular}{lccccccccc}
\multicolumn{1}{l}{Operator}&\multicolumn{9}{c}{Excited spin-\ohf\ and spin-\thf\ $\Lambda$-hyperons created/mixed}\\
 & $\Lambda(1405)$ & $\Lambda(1520)$ & $\Lambda(1600)$ & $\Lambda(1670)$ & $\Lambda(1690)$ & $\Lambda(1800)$ & $\Lambda(1810)$ & $\Lambda(?)$ & $\Lambda(1890)$ \\
 & $J^P=\ohf^-$ & $\thf^-$ & $\ohf^+$ & $\ohf^-$ & $\thf^-$  & $\ohf^-$  & $\ohf^+$  & $\thf^-$  & $\thf^+$  \\ \hline
$\mb{(1,\ 8)}_{-1}$ & 4 & 1 & 4 & 4 & 1 & 4 & 4 & 1 & 1\\
$\mb{(1,\ 8^{\prime})}_{-1}$ & - & 1 & - & - & 1 & - & - & 1 & 1\\
$\mb{(1,\ 16)}_{-1}$ & 1 & 5 & 1 & 1 & 5 & 1 & 1 & 5 & 5\\
$\mb{(0,\ 8)}_{-1}$ & 4 & 3 & 4 & 4 & 3 & 4 & 4 & 3 & 3\\
$\mb{(0,\ 8^{\prime})}_{-1}$ & - & 3 & - & - & 3 & - & - & 3 & 3\\
$\mb{(0,\ 16)}_{-1}$ & 3 & 7 & 3 & 3 & 7 & 3 & 3 & 7 & 7
\end{tabular}
\end{ruledtabular}
\caption{\label{elamos}For $m_x=m_y=\hat m,\;m_z=m_s$, the number of nondegenerate staggered states in the dominant continuum irreps of the listed excited $\Lambda$-hyperons.  The quark-model assignment for the $\Lambda(1810)$ is very tentative~\cite{PDBook}.}
\end{table}
\begin{table}
\begin{ruledtabular}
\begin{tabular}{lccccccccc}
\multicolumn{1}{l}{Operator}&\multicolumn{9}{c}{Excited spin-\ohf\ and spin-\thf\ $\Lambda_s$-hyperons created/mixed}\\
 & $\Lambda_s(?)$ & $\Lambda_s(?)$ & $\Lambda_s(?)$ & $\Lambda_s(?)$ & $\Lambda_s(?)$ & $\Lambda_s(?)$ & $\Lambda_s(?)$ & $\Lambda_s(?)$ & $\Lambda_s(?)$ \\
 & $J^P=\ohf^-$ & $\thf^-$ & $\ohf^+$ & $\ohf^-$ & $\thf^-$  & $\ohf^-$  & $\ohf^+$  & $\thf^-$  & $\thf^+$  \\ \hline
$\mb{(\ohf,\ 8)}_{-2}$ & 4 & 1 & 4 & 4 & 1 & 4 & 4 & 1 & 1\\
$\mb{(\ohf,\ 8^{\prime})}_{-2}$ & - & 1 & - & - & 1 & - & - & 1 & 1\\
$\mb{(\ohf,\ 16)}_{-2}$ & 1 & 5 & 1 & 1 & 5 & 1 & 1 & 5 & 5
\end{tabular}
\end{ruledtabular}
\caption{\label{estralamos}For $m_x=m_y=\hat m,\;m_z=m_s$, the number of nondegenerate staggered states in the dominant continuum irreps of the listed excited $\Lambda_s$-hyperons.  Note that the expected number of distinct continuum states for each $J^P$ is the same as for the $\Lambda$-hyperons listed in Table~\ref{elamos}, in accord with the observation that $\Lambda\leftrightarrow\Lambda_s$ under $\hat m\leftrightarrow m_s$.  Continuum energy levels with the same $J^P$ mix in the continuum limit.}
\end{table}
\begin{table}
\begin{ruledtabular}
\begin{tabular}{lccccccccc}
\multicolumn{1}{l}{Operator}&\multicolumn{9}{c}{Excited $\Sigma$-hyperons created/mixed}\\
 & $\Sigma(1620)$ & $\Sigma(1660)$ & $\Sigma(1670)$ & $\Sigma(1750)$ & $\Sigma(1775)$ & $\Sigma(?)$ & $\Sigma(1880)$ & $\Sigma(1915)$ & $\Sigma(?)$ \\
 & $J^P=\ohf^-$ & $\ohf^+$ & $\thf^-$ & $\ohf^-$ & $\fhf^-$  & $\thf^-$  & $\ohf^+$  & $\fhf^+$  & $\thf^+$  \\ \hline
$\mb{(1,\ 8)}_{-1}$ & 4 & 4 & 3 & 4 & 3 & 3 & 4 & 3 & 3\\
$\mb{(1,\ 8^{\prime})}_{-1}$ & - & - & 3 & - & 7 & 3 & - & 7 & 3\\
$\mb{(1,\ 16)}_{-1}$ & 3 & 3 & 7 & 3 & 10 & 7 & 3 & 10 & 7\\
$\mb{(0,\ 8)}_{-1}$ & 4 & 4 & 1 & 4 & 1 & 1 & 4 & 1 & 1\\
$\mb{(0,\ 8^{\prime})}_{-1}$ & - & - & 1 & - & 5 & 1 & - & 5 & 1\\
$\mb{(0,\ 16)}_{-1}$ & 1 & 1 & 5 & 1 & 6 & 5 & 1 & 6 & 5
\end{tabular}
\end{ruledtabular}
\caption{\label{esigos}For $m_x=m_y=\hat m,\;m_z=m_s$, the number of nondegenerate staggered states in the dominant continuum irreps of the listed excited $\Sigma$-hyperons.}
\end{table}
\begin{table}
\begin{ruledtabular}
\begin{tabular}{lccccccc}
\multicolumn{1}{l}{Operator}&\multicolumn{7}{c}{Excited $\Sigma^*$-hyperons and spin-\fhf\ $\Lambda$-hyperons created/mixed}\\
 & $\Sigma^*(?)$ & $\Sigma^*(?)$ & $\Sigma^*(?)$ & $\Sigma^*(?)$ & $\Sigma^*(2030)$ & $\Lambda(1820)$ & $\Lambda(1830)$ \\
 & $J^P=\ohf^-$ & $\thf^+$ & $\thf^-$ & $\fhf^+$ & $\shf^+$  & $\fhf^+$  & $\fhf^-$ \\ \hline
$\mb{(1,\ 8)}_{-1}$ & 4 & 3 & 3 & 3 & 7 & 1 & 1\\
$\mb{(1,\ 8^{\prime})}_{-1}$ & - & 3 & 3 & 7 & 7 & 5 & 5\\
$\mb{(1,\ 16)}_{-1}$ & 3 & 7 & 7 & 10 & 13 & 6 & 6\\
$\mb{(0,\ 8)}_{-1}$ & 4 & 1 & 1 & 1 & 5 & 3 & 3\\
$\mb{(0,\ 8^{\prime})}_{-1}$ & - & 1 & 1 & 5 & 5 & 7 & 7\\
$\mb{(0,\ 16)}_{-1}$ & 1 & 5 & 5 & 6 & 7 & 10 & 10
\end{tabular}
\end{ruledtabular}
\caption{\label{esigstarandlamos}For $m_x=m_y=\hat m,\;m_z=m_s$, the number of nondegenerate staggered states in the dominant continuum irreps of the listed excited $\Sigma^*$- and $\Lambda$-hyperons.}
\end{table}
\begin{table}
\begin{ruledtabular}
\begin{tabular}{lccccccccc}
\multicolumn{1}{l}{Operator}&\multicolumn{9}{c}{Excited $\Xi$-hyperons created/mixed}\\
 & $\Xi(?)$ & $\Xi(?)$ & $\Xi(?)$ & $\Xi(?)$ & $\Xi(1820)$ & $\Xi(?)$ & $\Xi(?)$ & $\Xi(?)$ & $\Xi(2030)$ \\
 & $J^P=\ohf^+$ & $\ohf^-$ & $\ohf^-$ & $\ohf^+$ & $\thf^-$  & $\thf^-$  & $\thf^+$  & $\fhf^-$  & $\fhf^+$  \\ \hline
$\mb{(\ohf,\ 8)}_{-2}$ & 4 & 4 & 4 & 4 & 3 & 3 & 3 & 3 & 3\\
$\mb{(\ohf,\ 8^{\prime})}_{-2}$ & - & - & - & - & 3 & 3 & 3 & 7 & 7\\
$\mb{(\ohf,\ 16)}_{-2}$ & 3 & 3 & 3 & 3 & 7 & 7 & 7 & 10 & 10
\end{tabular}
\end{ruledtabular}
\caption{\label{exios}For $m_x=m_y=\hat m,\;m_z=m_s$, the number of nondegenerate staggered states in the dominant continuum irreps of the listed excited $\Xi$-hyperons.  The quark-model assignments are very tentative.  The expected number of distinct continuum states for each $J^P$ is the same as for the $\Sigma$-hyperons listed in Table~\ref{esigos}, in accord with $\Sigma\leftrightarrow\Xi$ under $\hat m\leftrightarrow m_s$.}
\end{table}
\begin{table}
\begin{ruledtabular}
\begin{tabular}{lccccccc}
\multicolumn{1}{l}{Operator}&\multicolumn{7}{c}{Excited $\Xi^*$-hyperons and spin-\fhf\ $\Lambda_s$-hyperons created/mixed}\\
 & $\Xi^*(?)$ & $\Xi^*(?)$ & $\Xi^*(?)$ & $\Xi^*(?)$ & $\Xi^*(?)$ & $\Lambda_s(?)$ & $\Lambda_s(?)$ \\
 & $J^P=\ohf^-$ & $\thf^+$ & $\thf^-$ & $\fhf^+$ & $\shf^+$ & $\fhf^+$ & $\fhf^-$ \\ \hline
$\mb{(\ohf,\ 8)}_{-2}$ & 4 & 3 & 3 & 3 & 7 & 1 & 1 \\
$\mb{(\ohf,\ 8^{\prime})}_{-2}$ & - & 3 & 3 & 7 & 7 & 5 & 5\\
$\mb{(\ohf,\ 16)}_{-2}$ & 3 & 7 & 7 & 10 & 13 & 6 & 6
\end{tabular}
\end{ruledtabular}
\caption{\label{existaros}For $m_x=m_y=\hat m,\;m_z=m_s$, the number of nondegenerate staggered states in the dominant continuum irreps of the listed excited $\Xi^*$- and $\Lambda_s$-hyperons.  The expected number of distinct continuum states for each $J^P$ is the same as for the hyperons listed in Table~\ref{esigstarandlamos}, in accord with $\Sigma^*\leftrightarrow\Xi^*$ and $\Lambda\leftrightarrow\Lambda_s$ under $\hat m\leftrightarrow m_s$.}
\end{table}
\begin{table}
\begin{ruledtabular}
\begin{tabular}{lccccccccc}
\multicolumn{1}{l}{Operator}&\multicolumn{9}{c}{Excited $N_s$-hyperons created/mixed}\\ 
                        & $N_s(?)$ & $N_s(?)$ & $N_s(?)$ & $N_s(?)$ & $N_s(?)$ & $N_s(?)$ & $N_s(?)$ & $N_s(?)$ & $N_s(?)$ \\
                           & $J^P=\ohf^+$ & $\thf^-$ & $\ohf^-$ & $\ohf^-$ & $\fhf^-$  & $\fhf^+$  & $\thf^-$  & $\ohf^+$  & $\thf^+$  \\ \hline
$\mb{(0,\ 8)}_{-3}$          & 3         & 1         & 3         & 3         & 1         & 1         & 1         & 3         & 1         \\
$\mb{(0,\ 8^{\prime})}_{-3}$ & -         & 1         & -         & -         & 4         & 4         & 1         & -         & 1         \\
$\mb{(0,\ 16)}_{-3}$         & 1         & 4         & 1         & 1         & 5         & 5         & 4         & 1         & 4
\end{tabular}
\end{ruledtabular}
\caption{\label{estranucos}For $m_x=m_y=m_z=\hat m$, the number of nondegenerate staggered states in the dominant continuum irreps of the listed excited hyperons.  The expected number of distinct continuum states for each $J^P$ is the same as for the nucleons listed in Table~\ref{enucos}, in accord with $N\leftrightarrow N_s$ under $\hat m\leftrightarrow m_s$.}
\end{table}
\begin{table}
\begin{ruledtabular}
\begin{tabular}{lccccccc}
\multicolumn{1}{l}{Operator}&\multicolumn{7}{c}{Excited $\Omega^-$'s and $\Lambda_{ss}$'s created/mixed}\\
                         & $\Omega^-(?)$ & $\Omega^-(?)$ & $\Omega^-(?)$ & $\Omega^-(?)$ & $\Omega^-(?)$ & $\Lambda_{ss}(?)$ & $\Lambda_{ss}(?)$ \\
                         & $J^P=\thf^+$   & $\ohf^-$       & $\thf^-$       & $\fhf^+$       & $\shf^+$       & $\ohf^-$       & $\thf^-$ \\ \hline
$\mb{(0,\ 8)}_{-3}$          & 2              & 1              & 2              & 2              & 3              & 1              & - \\
$\mb{(0,\ 8^{\prime})}_{-3}$ & 2              & -              & 2              & 3              & 3              & -              & - \\
$\mb{(0,\ 16)}_{-3}$         & 3              & 2              & 3              & 5              & 7              & -              & 1
\end{tabular}
\end{ruledtabular}
\caption{\label{eomegos}For $m_x=m_y=m_z=\hat m$, the number of nondegenerate staggered states in the dominant continuum irreps of the listed excited $\Omega^-$'s and $\Lambda_{ss}$'s.  The expected number of distinct continuum states for each $J^P$ is the same as for the $\Delta$'s and $\Lambda_u$'s listed in Table~\ref{edelos}, in accord with $\Delta\leftrightarrow\Omega^-$ and $\Lambda_u\leftrightarrow\Lambda_{ss}$ under $\hat m\leftrightarrow m_s$.}
\end{table}
Several aspects of these results deserve further comment.  First, the spectra generically include both positive and negative parity states.  This situation does not present any problem in practice because one can fit the associated oscillations in the correlators to extract the masses of both.  For example, consider the excited states created by operators in the irreps of Table~\ref{best}.  For degenerate valence quarks equal to $\hat m$, the spectrum of operators transforming in \mb{(8_M,\ 8^{\prime})} contains a single $\Delta(1232)$, a single $J^P=\thf^-$ $\Lambda_u$, three states that correspond in the continuum limit to the dominant representation of the $N(1520)$, a single $\Delta(1600)$ representation, and representations corresponding to various heavier excited nucleons and $\Delta$'s.  Because the $\Lambda_u$ is a negative parity state, its contribution to the correlator will oscillate and can be removed even if its mass is close to that of the $\Delta(1232)$.  The next nearest state is the $N(1520)$, separated by nearly 300 MeV.  Similar considerations apply to the spectra of the other operators listed in Table~\ref{best}.  We are therefore reassured that contamination from excited states will not be problematic when extracting the masses of the ground state nucleon and decuplet.  

Second, all the states of a given parity created by a given operator mix for nonzero lattice spacing.  For example, the spectrum of operators in the \mb{(10_S,\ 16)} includes not only the nucleons of Table~\ref{enucos}, but also the $\Delta$'s and spin-$\thf^-$ $\Lambda_u$ overlapped in accord with Table~\ref{edelos}.  The state corresponding to the $N(1440)$ mixes with those corresponding to, e.g., the $N(1680)$, the $N(1710)$, and the $N(1720)$, but not with the $\Lambda_u$.  

Finally, Tables~\ref{enucos}, \ref{edelos}, \ref{elamos}, \ref{estralamos}, \ref{esigos}, \ref{esigstarandlamos}, \ref{exios}, \ref{existaros}, \ref{estranucos}, and \ref{eomegos} allow us to consider how best to extract the spectrum of excited states.  Calculating the excited spectrum, an already formidable task~\cite{Basak:2006ww}, promises to be more so when using staggered valence quarks.

\bibliography{ops}

\end{document}